\documentclass[11pt]{article}
\usepackage{cite}
\usepackage{amsmath,amsfonts,amssymb}
\usepackage[small,bf,hang]{caption}
\usepackage{slashed}
\usepackage{color}

\allowdisplaybreaks

\def\hybrid{
        \topmargin -20pt
        \oddsidemargin 0pt
        \headheight 0pt \headsep 0pt
        \textwidth 6.25in 
        \textheight 9.5in 
        \marginparwidth .875in
        \parskip 5pt plus 1pt \jot = 1.5ex}

\hybrid

 \csname
@addtoreset\endcsname{equation}{section}

\def\moth{\mathsurround=0pt}
\newdimen\zo \zo=0pt

\def\tick{\leaders\hrule height 0.5ex depth 0pt \hskip 0.5pt}
\def\upboxfill{$\moth \setbox\zo\hbox{\tick}%
  \hskip 3pt\hbox to 0pt{$\tick$\hss}\hrulefill \hbox to 7.5pt{$\tick$\hss}$}

\def\dtick{\leaders\hrule height .34pt depth 0.5ex \hskip 0.5pt}
\def\downboxfill{$\moth \setbox\zo\hbox{\dtick}%
  \hskip 2pt\hbox to 0pt{$\dtick$\hss}\hrulefill \hbox to 2pt{$\dtick$\hss}$}

\def\bec{\begin{center}}
\def\ec{\end{center}}

\def\be{\begin{equation}}
\def\ee{\end{equation}}
\def\bea{\begin{eqnarray}}
\def\eea{\end{eqnarray}}
\def\ba{\begin{array}}
\def\ea{\end{array}}

\begin{document}

\begin{titlepage}
\rightline{}
\rightline{July  2022}
\rightline{HU-EP-22/17-RTG}
\begin{center}
\vskip 1.5cm
 {\Large \bf{Cosmological Perturbations in Double Field Theory}}
\vskip 1.7cm

{\large\bf {Olaf Hohm and Allison F.~Pinto}}
\vskip 1.6cm

{\it  Institute for Physics, Humboldt University Berlin,\\
 Zum Gro\ss en Windkanal 6, D-12489 Berlin, Germany}\\
\vskip .1cm

\vskip .2cm

ohohm@physik.hu-berlin.de, apinto@physik.hu-berlin.de

\end{center}

\bigskip\bigskip\bigskip 
\begin{center} 
\textbf{Abstract}

\end{center} 
\begin{quote}

We explore perturbative double field theory about time-dependent (cosmological) backgrounds to cubic order. 
To this order the theory is consistent in a weakly constrained sense, so that for a  toroidal geometry it encodes both 
momentum and genuine 
winding modes. We give a  self-contained discussion of the consistency problems and their resolution, 
including the role of cocycle factors and  the $O(d,d,\mathbb{Z})$ duality. 
Finally, as a first step toward the computation of cosmological 
correlation functions, we propose a generalized scalar-vector-tensor decomposition and use it to 
construct gauge invariant generalized Bardeen variables. 
Compared to standard cosmology there are fewer  tensor modes but more vector and scalar modes.

\end{quote} 
\vfill
\setcounter{footnote}{0}
\end{titlepage}

\tableofcontents


\section{Introduction}

Cosmology requires, ultimately,  a gravitational theory beyond Einstein's  general relativity in order to resolve the Big Bang 
singularity. As a consistent theory of quantum gravity, string theory is a promising candidate for such a theory. 
While it has turned out to be very challenging to obtain predictions from  string theory in the realm of particle physics, 
in cosmology the situation may be better, given that here effects of fundamental physics 
at microscopic scales may be amplified to very large scales. 
Since the early days of string theory there have been striking ideas of how to exploit some of the 
unique characteristics of string theory, such as  dualities, in cosmological scenarios \cite{Brandenberger:1988aj,Tseytlin:1991xk,Veneziano:1991ek,Gasperini:1996fu,Gasperini:2002bn} (see \cite{Brandenberger:2018xwl,Brandenberger:2018wbg,Bernardo:2020nol,Bernardo:2020bpa,Nunez:2020hxx,Brandenberger:2021zib} 
for some recent developments).

The simplest duality emerges directly in string cosmologies with vanishing spatial curvature (that luckily 
appears to be a feature  of the universe we live in). In this case, the dynamics of the Friedmann-Robertson-Walker (FRW) 
backgrounds are encoded   in the dimensional reduction of the string target space actions to one dimension (cosmic time), 
for which classical string theory exhibits the enhanced symmetry group $O(d,d,\mathbb{R})$  to all orders in 
the  inverse string tension $\alpha'$ \cite{Meissner:1991zj,Meissner:1991ge,Sen:1991zi}, with $d$ being  the number 
of spatial dimensions.  This group contains 
the duality transformation $a(t)\rightarrow \frac{1}{a(t)}$ of the cosmological scale factor  (`scale factor duality') that is at the heart of some 
of the string cosmology proposals \cite{Brandenberger:1988aj,Tseytlin:1991xk,Veneziano:1991ek}.

In modern cosmology the small deviations or fluctuations around FRW spacetimes are equally important, 
as these relate to the temperature fluctuations of the cosmic microwave background (CMB), see 
\cite{Mukhanov:1990me} for a review. 
In order to include  such fluctuations it is no longer sufficient to employ the dimensionally reduced theory; rather, 
in cosmological perturbation theory 
the fundamental  fields are  allowed to depend on all spatial coordinates, in which case the  $O(d,d,\mathbb{R})$
symmetry is generically broken to $GL(d,\mathbb{R})$. 
However, if the spatial geometry of the FRW spacetime contains a torus (as assumed in the string-gas cosmology 
proposal of \cite{Brandenberger:1988aj}) then in string theory there is still an $O(d,d,\mathbb{Z})$ duality invariance. 
This is due to the appearance of new states or fields that are not present in standard gravity, the so-called winding modes
of the string, which under $O(d,d,\mathbb{Z})$ are mapped to the more familiar Kaluza-Klein modes. 
It is of great interest to explore whether these modes, or more general massive string modes, assuming they were  excited 
in the very early universe, can leave an imprint on the CMB.

In this paper we explore some general aspects of the corresponding  field theory, 
using double field theory \cite{Siegel:1993th,Hull:2009mi,Hohm:2010jy,Hohm:2010pp}. 
Here the fields depend in addition to cosmic time $t$ on doubled spatial coordinates ${\bf X}=(x,\tilde{x})$, 
where upon Fourier expansion the dependence on the dual coordinates $\tilde{x}$ leads to the infinite 
tower of winding modes. More precisely, in string theory the fields are subject to the level-matching constraint, 
which requires that for a given Fourier mode the momenta ${\bf k}$ and the winding vector ${\bf w}$ are orthogonal, 
i.e., ${\bf k}\cdot{\bf w}=0$. The implementation of this constraint in a consistent theory (other than the full string theory) 
is a difficult and so far unsolved problem,\footnote{However, on general grounds this theory has to exist \cite{Sen:2016qap,Arvanitakis:2021ecw}.} but on flat backgrounds to cubic order such a  double field theory 
 was derived by Hull and Zwiebach in \cite{Hull:2009mi} and is entirely consistent to this order. 
This theory is usually referred to as `weakly constrained'. Most of the subsequent work on double field theory 
was done on the `strongly constrained' subsector for which a stronger version of the level-matching constraint 
is assumed, which in turn implies that the doubling of coordinates is only formal. (However,  this theory 
still makes the $O(d,d,\mathbb{R})$ symmetry 
arising upon dimensional reduction manifest \textit{before}  reduction and thus has found numerous applications, 
see \cite{Hohm:2013bwa} for a review.)

Since in cosmological perturbation theory much information is already contained in the quadratic theory around 
FRW backgrounds, which encodes the two-point functions, or the cubic theory, which would be required in order to 
compute the so far unobserved three-point functions possibly encoding non-Gaussianity \cite{Maldacena:2002vr}, 
we will here revisit true or weakly constrained double 
field theory to cubic order, 
but allowing the background to be a time-dependent FRW spacetime whose spatial geometry is a torus. 
It is a notoriously difficult problem to treat time-dependent backgrounds in string theory, and therefore we will not attempt 
to derive this double field theory directly from closed string field theory. Instead we use the short cut of starting from the 
strongly constrained double field theory which is known in a \textit{background independent} form \cite{Siegel:1993th,Hohm:2010jy}, so that 
we may expand about an arbitrary background solution, including time-dependent ones. 
Following Hull and Zwiebach we may then reinterpret the resulting theory to cubic order (and so far to cubic order only) 
as a weakly constrained theory. More precisely, this holds up to so-called cocycle factors, which are claimed 
to be necessary for consistency to quartic and higher orders \cite{Hata:1986mz,Maeno:1989uc,Kugo:1992md}, 
and which have not been included in  \cite{Hull:2009mi}. 
For this issue the time dependence of the background does not seem relevant, and we use the opportunity 
to give a complete discussion of the consistency of the cubic theory including the cocycles, 
following the treatment by Kugo and Zwiebach \cite{Kugo:1992md}. 
While we thus do not provide a first-principle derivation of this theory, the latter  is  consistent to 
cubic order and has all features expected of string theory for such backgrounds.

Although the above program seems a straightforward extension of double field theory on flat space, 
it turns out that the structure of the theory is significantly more involved for time-dependent backgrounds. 
Ultimately these complications trace back to the fact that in string theory on FRW backgrounds the symmetry 
group is reduced compared to flat space. More precisely, there is the duality group 
$O(d,d,\mathbb{Z})$ that maps different backgrounds into each other in such a way that string  theories 
on such backgrounds are equivalent. The subgroup that leaves the background genuinely invariant 
is then a symmetry group of string theory. For a flat metric  this group is $O(d)\times O(d)$, 
but for non-trivial  FRW backgrounds it is reduced to the diagonal subgroup hereof, 
 \be
  (O(d)\times O(d))_{\rm diag} \ \subset \ O(d)\times O(d) \ \subset \  O(d,d)\;. 
  \ee
Concretely, the complication is that while  perturbative double field theory on any background features a `double copy' type 
factorization into unbarred/barred indices \cite{Hohm:2011dz}, only on flat space does this imply two independent 
rotational symmetries. 
In contrast, for FRW backgrounds a tensor $L_{a}{}^{\bar{b}}$ enters the theory that is proportional to 
the Hubble parameter $H=\frac{\dot{a}}{a}$ and which  connects unbarred and barred indices. 
Consequently, the double field theory action allows for contractions of the derivative operators $D_a$ and $D_{\bar{a}}$ 
that are linearly independent combinations of momentum and winding operators. 
Concretely, for a flat FRW background with a single scale factor $a(t)$, these operators 
are given in terms of the dual momentum derivatives $\partial$ and 
winding derivatives $\tilde{\partial}$ by 
 \be
   D \ \propto \ a^{-1}(t) \partial - a(t) \tilde{\partial} \;, \qquad \bar{D} \ \propto \  a^{-1}(t) \partial +a(t) \tilde{\partial}\;. 
 \ee
In contrast to flat space, where $a\equiv 1$, there is no doubled spatial rotation invariance, 
which in turn complicates 
the decomposition of fields into irreducible vector and tensor components 
subject to divergence-free and other  constraints. 
Here we propose a  scalar-vector-tensor (SVT) decomposition that has two desirable features: 
First, it allows for  the definition of gauge invariant field variables, generalizing those introduced by Bardeen in conventional cosmology \cite{Bardeen:1980kt}. Second, in terms of these variables the quadratic action completely  decouples 
between  tensor, vector and scalar modes.
However, this SVT decomposition is significantly more involved than in standard cosmology since, 
owing to the presence of twice as many derivatives, one may impose more constraints, say on the divergence of a tensor.  
Consequently, a decomposition into fully irreducible objects (that  cannot be subjected to further constraints)  
leads to more vector and scalar modes than 
in standard cosmology. It is tempting  to speculate about the possible observability of these extra modes.

The rest of this paper is organized as follows. In sec.~2 we give a brief review of the background independent 
double field theory, and we expand it about an arbitrary time-dependent background to cubic order. 
In particular, we display the gauge symmetries  to this order and verify invariance. In sec.~3 we discuss the duality 
properties of time-dependent backgrounds in double field theory and then specialize to a particular class of 
FRW-type backgrounds and discuss their invariance group. In the final subsection we explain  that the theory 
to cubic order is consistent in a weakly constrained sense, and we introduce the cocyle factors that appear to be 
necessary in string theory. In sec.~4 we introduce a scalar-vector-tensor decomposition and use it to 
define gauge invariant variables. 
These technical results should be the basis, for instance, for the computation of cosmological correlation functions, 
which we leave for future work. 
We briefly discuss this and other aspects in a conclusion section.

\section{Double field theory in time-dependent backgrounds}
In this section we give a brief review of double field theory and then expand its action  around a general time-dependent background. 
We display the action quadratic in fluctuations and discuss its gauge invariance together with its canonical formulation. 
Finally, we give the complete cubic action. 

\subsection{Generalities}
The double field theory action can be written in terms of the generalized metric and the dilaton on a doubled space. The canonical formulation  was derived in \cite{Naseer:2015fba}, based on earlier results in \cite{Hohm:2013nja}, via splitting the doubled coordinates and indices into temporal and spatial components and imposing the condition that all fields are independent of dual time. This results in a theory with fields: 
 \begin{equation}
 {\cal H}_{MN}\,,\quad  \Phi\, , \quad n\, , \; \quad {\cal N}^M\;, 
 \end{equation}
where ${\cal H}_{MN}$ denotes the spatial generalized metric, satisfying $\mathcal{H}_{MK}\mathcal{H}^{KN}= \delta_{M}{}^N$, $\Phi$ is the duality invariant dilaton, $n$ denotes the lapse function ensuring time reparametrization invariance, 
and ${\cal N}^M$ denotes the doubled shift vector. The action is  formulated on a $(1+2d)$-dimensional space and reads 
\begin{equation} 
\label{RealStartingAction}
S = \int dt \int d^{2d} {\bf X} \,n\, e^{-2\Phi}  \bigg( - 4 (D_t \Phi )^2 - \frac{1}{8} D_t \mathcal{H}_{MN} D_{t} \mathcal{H}^{MN} +  \mathcal{R}(\Phi,\mathcal{H}_{MN}  ) \bigg)  \,. 
\end{equation} 
Here doubled spatial coordinates are denoted by ${\bf X}^M=(\tilde{x}_i, x^i)$, with $i=1,...,d$ and $M=1,...,2d$, and all fields 
depend on $(t, {\bf X}^M)$. 
The covariant derivatives are defined as  
 \bea
  D_t \equiv \frac{1}{n}\left(\partial_t - \mathcal{L}_\mathcal{N}\right)\;,  
 \eea
 where $\mathcal{L}_\mathcal{N}$ is the generalized Lie derivative with respect to the generalized shift vector $\mathcal{N}^M$: 
 \begin{equation}
  \begin{split}
   {\cal L}_{{\cal N}}\Phi \ &= \ {\cal N}^{M}\partial_{M} \Phi -\frac{1}{2}\partial_M{\cal N}^{M}\;, \\
   {\cal L}_{\cal N}{\cal H}_{MN} \ &= \ {\cal N}^{K}\partial_{K}{\cal H}_{MN} + K_{M}{}^{K}({\cal N}){\cal H}_{KN} + K_{N}{}^{K}({\cal N}) {\cal H}_{KN}\;. 
  \end{split}
 \end{equation} 
Here we defined $K_{MN}(\xi)=2\partial_{[M}\xi_{N]}$, where $O(d,d)$ indices are contracted by the $O(d,d)$ metric 
\begin{equation}\label{Oddddmetric}
\eta_{MN}  \ =  \ \begin{pmatrix} 0 & \delta^i{}_j  \\ \delta_i{}^j & 0 \end{pmatrix}  \,. 
\end{equation}  
The last term in the action (\ref{RealStartingAction}) uses the generalized curvature scalar 
\begin{align}
\label{eq:genR}
    \begin{split}
        \mathcal{R}(\Phi,\mathcal{H}_{MN} ) \ &\equiv \  4 \mathcal{H}^{MN} \partial_M \partial_N \Phi  - \partial_M \partial_N \mathcal{H}^{MN} - 4 \mathcal{H}^{MN} \partial_M\Phi \partial_N \Phi + 4 \partial_M \mathcal{H}^{MN} \partial_N \Phi  \\
        & \;\; + \frac{1}{8} \mathcal{H}^{MN} \partial_M \mathcal{H}^{KL} \partial_N \mathcal{H}_{KL}
        - \frac{1}{2} \mathcal{H}^{MN} \partial_M \mathcal{H}^{KL} \partial_K \mathcal{H}_{NL} \;. 
    \end{split}
\end{align}
For now we assume all fields to be subject to the strong version of the constraint 
\begin{equation}
\label{weakconstraint}
\eta^{MN}\partial_M \partial_N f = 2 \tilde{\partial}^i \partial_i f  = 0 \,, 
\end{equation}
meaning that also  terms of the form $\partial^Mf\, \partial_Mg$ are set to zero.  

In the following we will employ a frame formalism for the internal generalized metric 
\cite{Siegel:1993th,Hohm:2010xe}. 
We introduce a generalized frame or 
vielbein $E_A{}^M$, with inverse $E_M{}^{A}$, satisfying 
$E_A{}^M E_{M}{}^{B}=\delta_A{}^{B}$, from which the generalized metric can be constructed via
 \bea
  \mathcal{H}_{MN} \ = \ E_M{}^A E_N{}^B S_{AB}\;, 
 \eea
where $S_{AB}$ denotes a positive-definite (tangent space) metric to be given momentarily. The flat indices split as $A=(a,\bar{a})$. 
The frame field is subject to the constraint that the ``flattened" version of the $O(d,d)$ metric (\ref{Oddddmetric}) is block-diagonal:  
 \be\label{flatetaisG}
  {\cal G}_{AB} \ \equiv \ E_A{}^{M} E_{B}{}^{N} \eta_{MN} \ = \ \begin{pmatrix} {\cal G}_{ab} & 0 \\ 0 & {\cal G}_{\bar{a}\bar{b}} \end{pmatrix}\;, 
 \ee
with no further constraints on the (generally spacetime dependent) metrics ${\cal G}_{ab}$ and ${\cal G}_{\bar{a}\bar{b}}$.  
Thus, the local frame transformations comprise the group $GL(d,\mathbb{R})\times GL(d,\mathbb{R})$.   
The metric ${\cal G}_{AB}$, which is used to raise and lower flat indices, has signature $(d,d)$, and so we can assume without loss of generality that ${\cal G}_{ab}$ is negative-definite and that ${\cal G}_{\bar{a}\bar{b}}$ 
is positive-definite. Then, the metric $S_{AB}$ is defined as 
 \be\label{SMatrix}
  S_{AB} \ = \ \begin{pmatrix} -{\cal G}_{ab} & 0 \\ 0 & {\cal G}_{\bar{a}\bar{b}} \end{pmatrix}\;. 
 \ee

Finally, we give the symmetries of 
the action, which  is invariant under $O(d,d)$ transformations and generalized diffeomorphisms. The transformation rules can be inferred from \cite{Hohm:2013nja}. 
The gauge parameters  are ${\xi}^0$, $\tilde{\xi}_0$, $\xi^M$, and $\Lambda_A{}^{B}$, all depending on coordinates $(t,{\bf X}^M)$, and act 
infinitesimally via 
\begin{align}
\begin{split}
\label{gauge_transfr_reparametrized}
\delta E_A{}^M &= \mathcal{L}_{\xi} E_A{}^M + n\xi^0 D_t E_A{}^M  +\Lambda_{A}{}^{B} E_{B}{}^{M}\,,  \\ 
\delta n &= n \xi^0D_t n + n^2 D_t \xi^0  + \xi^M \partial_M n \,, \\
\delta \mathcal{N}^M &= \partial^M \tilde{\xi}_0 + \partial_t \xi^M - n^2 \mathcal{H}^{MN}\partial_N \xi^0 + \mathcal{L}_{\xi} \mathcal{N}^M   \,,  \\ 
\delta\Phi & = n\xi^0 D_t \Phi + \mathcal{L}_\xi \Phi  \,. 
\end{split}
\end{align}  

\subsection{Quadratic theory}
Now let us expand around a background that is purely time-dependent as follows:
\begin{flalign}
E_A{}^M(t,{\bf X}) \ &=  \ \bar{E}_A{}^M(t) - h_A{}^B (t,{\bf X})\bar{E}_B{}^M  (t) \,, \\ 
\Phi(t,{\bf X}) \ &= \ \bar{\Phi}(t)+ \varphi(t,{\bf X})\,,\\
n(t,{\bf X}) \ &= \ \bar{n}(t) (1 + \phi(t,{\bf X}) ) \,,\\
\mathcal{N}^M(t,{\bf X}) \ &= \ \bar{n}(t) \mathcal{A}^M (t,{\bf X})\,. 
\end{flalign}  
In the following we will omit the bar on $E_A{}^M(t)$, $\Phi$ and $n$ as these will exclusively refer to the background quantities. 
The background tangent space metric $\bar{\cal G}_{AB}$ constructed as in (\ref{flatetaisG}) will be 
used to raise and lower flat indices, while the background ${E}_{A}{}^{M}$ and its inverse will be used to flatten and unflatten indices. 
 Moreover, we assume that a gauge has been chosen for the (background) frame transformations for which 
 $\bar{\mathcal{G}}_{AB}$ is constant and hence does not depend on time. 
The corresponding background version of the metric (\ref{SMatrix}) is then also constant. 
This has the important advantage that we can freely raise and lower flat indices under time derivatives.

Finally, we note that the first-order frame transformations allow one to fix a gauge with $h_{ab}=h_{\bar{a}\bar{b}}=0$, so that the independent fluctuation is given by 
 \be
  h_{a\bar{b}} \ = \ -h_{\bar{b}a}\;, 
 \ee
where the last equality follows from the constraint that (\ref{flatetaisG}) is block-diagonal, i.e., ${\cal G}_{a\bar{b}}=0$. 
We have thus completely  fixed the first-order frame transformations.  

\subsubsection*{Background equations}
Inserting the above expansion into the action, one obtains to leading order an action for the purely time-dependent background fields encoding their dynamics: 
\begin{equation}
S_0 = \int dt \int d^{2d} {\bf X}  \, n ^{-1} \, e^{-2\Phi}\bigg( -4 \dot{\Phi}^2 -  \frac{1}{8}{\rm tr}\big( \dot{S}^2\big) \bigg) \,, 
\end{equation} 
where we employ matrix notation, with $S^M{}_N$ the background generalized metric with one index raised, and the dot denotes the time derivative. 
The field equations read 
\begin{equation}
\label{eomS}
\begin{split}
\ddot{S} + S \dot{S}^2 - 2\bigg(\dot{\Phi}+ \frac{1}{2} \partial_t \log{n } \bigg)  \dot{S} &= 0  \,, \\
-4 \ddot{\Phi} +4 \dot{\Phi}^2+4 \dot{\Phi}\partial_t \log{n }  - \frac{1}{8} {\rm tr}\big( \dot{S}^2 \big) &=0 \,, \\ 
4\dot{\Phi}^2 + \frac{1}{8} {\rm tr} \big(\dot{S}^2\big) &=0 \,.
\end{split}
\end{equation}
It will be convenient to express these equations explicitly in terms of the background frame field. 
We define 
\be
L_{A}{}^{B}   \ \equiv \ \frac{1}{n}\, \partial_t {E}_{A}{}^M {E}_{M}{}^B\;, 
\ee
in terms of which the equations of motion are \begin{flalign}
\label{eomLdot}
 \frac{1}{n}\dot{L}_{a}{}^{\bar{b}} - L_a{}^{c} L_c{}^{\bar{b}} +L_a{}^{\bar{c}} L_{\bar{c}}{}^{\bar{b}}- \frac{2}{n} \dot{\Phi}  L_{a}{}^{\bar{b}}  \ &= \ 0 \,, \\
 \label{eom phi original}
-4 \ddot{\Phi} + 4\dot{\Phi}^2 +4 \dot{\Phi}\partial_t \log{n } + n^2 \, L_a{}^{\bar{b}}L_{\bar{b}}{}^a   \ &=  \ 0 \,, \\
\label{eomdotPhi LL}
 4\dot{\Phi}^2  - n^2 \, L_a{}^{\bar{b}}L_{\bar{b}}{}^a  \ & = \ 0 \,.
\end{flalign}

It is instructive to rewrite these equations in terms of derivatives that are covariant under background frame transformations with parameter $\bar{\Lambda}_A{}^B$
and time reparametrization with parameter $\bar{\xi}^0$. These transformations act on $L_{A}{}^{B}$, the lapse function $n$ and generic vectors ${\cal V}_a$ and ${\cal V}_{\bar{a}}$ 
as 
\begin{align}\begin{split}
\bar{\delta} L_A{}^B &= \bar{\xi}^0 \partial_t  L_A{}^B+\frac{1}{n} \partial_t \bar{\Lambda}_A{}^B + \bar{\Lambda}_A{}^C L_C{}^B -  \bar{\Lambda}_C{}^B L_A{} ^C\;, \\
\bar{\delta} n  & = \partial_t (\bar{\xi}^0 n ) \,,\\
\bar{\delta}\mathcal{V}_a & = \bar{\xi}^0 \partial_t \mathcal{V}_a + \bar{\Lambda}_a{}^b\mathcal{V}_b  \,, \\
\bar{\delta} \mathcal{V}_{\bar{a}} & = \bar{\xi}^0 \partial_t \mathcal{V}_{\bar{a}} + \bar{\Lambda}_{\bar{a}} {}^{\bar{b}}\mathcal{V}_{\bar{b}} 
 \,. 
\end{split}\end{align} 
All other fields transform as scalars under $\bar{\xi}^0$ and as tensors under $\bar{\Lambda}_A{}^B$. 
The first line implies that $L_{a}{}^{b}$ and $L_{\bar{a}}{}^{\bar{b}}$ transform as connections under background frame 
transformations, while $L_{a}{}^{\bar{b}}$ transforms as a tensor. 
We then have the covariant derivatives 
\begin{align}
\begin{split}
\label{def_bkgrd_nablabar} 
{\nabla}_t \mathcal{V}_a \ &= \ \frac{1}{n } \partial_t \mathcal{V}_a - L_a{}^b \mathcal{V}_b \,, \\ 
{\nabla}_t \mathcal{V}_{\bar{a}} \ & =  \ \frac{1}{n } \partial_t \mathcal{V}_{\bar{a}} - L_{\bar{a}}{}^{\bar{b}} \mathcal{V}_{\bar{b}}  \,, 
\end{split}
\end{align}  
which transform covariantly in that 
\begin{align}
\begin{split}
\bar{\delta}\left(\nabla_t \mathcal{V}_a\right) \ & = \ \bar{\xi}^0 \partial_t \big(\nabla_t \mathcal{V}_a\big) + \bar{\Lambda}_a{}^b\nabla_t \mathcal{V}_b \,, \\ 
\bar{\delta}\left(\nabla_t \mathcal{V}_{\bar{a}}\right) \ & =  \ \bar{\xi}^0 \partial_t \big(\nabla_t\mathcal{V}_{\bar{a}}\big) 
+ \bar{\Lambda}_{\bar{a}} {}^{\bar{b}}\nabla_t\mathcal{V}_{\bar{b}} \,. 
\end{split}  
\end{align}
(\ref{eomLdot}) can now be written more compactly as
\begin{equation}
\label{eomLdotsimple}
{\nabla}_t L_a{}^{\bar{b}} - \frac{2}{n } \dot{\Phi} L_{a}{}^{\bar{b}}  \ =  \ 0 \,, 
\end{equation}
or, equivalently, 
 \be
 \label{nabla t L e zero}
  \nabla_t\left(e^{-2\Phi}L_{a\bar{b}}\right) \ = \ 0\;.  
 \ee 
In addition, upon adding (\ref{eom phi original}) and (\ref{eomdotPhi LL})
one obtains the useful equation 
\begin{equation}
\label{eomddotPhi} 
\ddot{\Phi} - 2 \dot{\Phi}^2  -\dot{\Phi}\,   \partial_t \log{n }  \ = \ 0  \,, 
\end{equation}
or, equivalently, 
 \be\label{simpdialtoneq}
  \partial_t\left(e^{-2\Phi}n^{-1}\dot{\Phi}\right) \ = \ 0\;. 
 \ee

\subsubsection*{Quadratic fluctuations}

Next, we assume that the background equations  are satisfied, so that the terms linear in fluctuations drop out. 
The action for the quadratic fluctuations is given by 
\begin{align}
\begin{split}
\label{SimpleActionQuad}
S^{(2)}=& \int dt \int d^{2d}{\bf X} \,\, \mathcal{L} \,, 
\end{split}
\end{align}
where 
\begin{equation}
\begin{split} \label{SimpleActionQuadLagrangian}
\mathcal{L} = n\,  e^{-2\Phi} \bigg\{ &
-  4 \Gamma^2+ 2 \phi \bigg(  4\frac{\dot{\Phi}}{n}  \Gamma
 - L^{a\bar{b}}  \omega_{a\bar{b}} \bigg) 
-\omega_{a\bar{b}} \omega^{a\bar{b}}  + \frac{1}{2} ( \omega_{ab} \omega^{ab}+ \omega_{\bar{a}\bar{b}} \omega^{\bar{a}\bar{b}}) \\
&- \frac{1}{2} \big( K_{ab}K^{ab} + K_{\bar{a}\bar{b}} K^{\bar{a}\bar{b}} \big)  + 4 L^{a\bar{b}} ( \varphi K_{a\bar{b}} +  \Gamma h_{a\bar{b}} ) 
 +\mathcal{V}^{(2)}(\phi, \varphi, h)\bigg\} \,, 
\end{split}
\end{equation}
and we defined 
\begin{align}
\begin{split}
\Gamma &\equiv  \nabla_t \varphi + \frac{1}{2} \partial_M \mathcal{A}^M\,,\\\
\omega_{a\bar{b}}  &\equiv \nabla_t h_{a\bar{b}} - K_{a\bar{b}} \,,\\
\omega^{ab} &\equiv2h^{[a}{}_{\bar{c}}L^{b]\bar{c}} + K^{ab} \,, \\
\omega^{\bar{a}\bar{b}} &\equiv 2h^{c[\bar{a}}L_{c}{}^{\bar{b}]} + K^{\bar{a}\bar{b}}   \,, 
\end{split}
\end{align}
with 
 \be
  K_{AB}\equiv K_{AB}({\cal A}) = D_{A}{\cal A}_{B}-D_{B}{\cal A}_{A}\;, 
 \ee
and 
 \be
  D_A \ \equiv \ \bar{E}_{A}{}^{M}\partial_M 
  \;.  
 \ee
 $K_{AB}$ satisfies the identities:
 \begin{equation}
 \label{K Bianchi}
D_{A}K_{BC} + D_{B} K_{CA} + D_{C}K_{AB}=0 \,,
\end{equation}
\begin{equation}
\label{Aweakconstraint}
D^B K_{AB}  - D_{A} D^B \mathcal{A}_B = 0 \,,
\end{equation}
the latter being a consequence of the strong constraint (\ref{weakconstraint}), which in terms of this flattened derivative takes the form
\begin{equation}
\Delta \equiv -2 D^a D_a = 2 D^{\bar{a}} D_{\bar{a}} \,. 
\end{equation}
 Also note that the flattened spatial derivatives and the covariant (time) derivative satisfy the commutation relations: 
\begin{flalign}
\label{Commutator a}
\big[D_a , \nabla_t  \big] \mathcal{V}_{B} & =- L_{a}{}^{\bar{c}} D_{\bar{c}} \mathcal{V}_{B} \,, \\ 
\label{Commutator bar}
\big[D_{\bar{a}} , \nabla_t  \big] \mathcal{V}_{B} & =- L_{\bar{a}}{}^{c} D_{c} \mathcal{V}_{B} \,.
\end{flalign}
In the action we collected the terms with only spatial derivatives into ${\cal V}^{(2)}$, defined as  
\begin{align}
\begin{split} 
 \mathcal{V}^{(2)}(\phi, \varphi, h_{a\bar{b}}) \  = \ \; & 
8 D^a \phi D_a \varphi - 8   D^a \varphi D_a \varphi 
-8 D_a  \varphi D_{\bar{b}} h^{a\bar{b}}  
+ 4 D_a \phi D_{\bar{b}} h^{a\bar{b}} 
-2 D^{a} h^{b\bar{c}} D_{a} h_{b\bar{c}} \\ & \, 
+ 2D^c h^{a\bar{b}} D_a h_{c\bar{b}} 
- 2 D^{\bar{c}} h^{a\bar{b}} D_{\bar{b}} h_{a\bar{c}} \,. 
\end{split}
\end{align}
This can be rewritten in terms of 
 \be
  \varphi_{\pm} \ \equiv \ \varphi\pm\frac{1}{2}\phi\;,  
 \ee
which yields 
\begin{align}
\begin{split} 
 \mathcal{V}^{(2)}(\phi, \varphi_{-}, h_{a\bar{b}}) \  = \ \; & 
2 D^a \phi D_a \phi - 8   D^a \varphi_{-} D_a \varphi_{-} 
-8 D_a  \varphi_{-} D_{\bar{b}} h^{a\bar{b}}  
-2 D^{a} h^{b\bar{c}} D_{a} h_{b\bar{c}} \\ & \, 
+ 2D^c h^{a\bar{b}} D_a h_{c\bar{b}} 
- 2 D^{\bar{c}} h^{a\bar{b}} D_{\bar{b}} h_{a\bar{c}} \,. 
\end{split}
\end{align}

We next turn to the gauge invariance of the quadratic action, which for time-dependent backgrounds is 
quite subtle even for the free theory.  Under linearized gauge transformations one finds 
\begin{align}\label{quadrgauge1234}
\begin{split}
\delta^{(0)} h_{a\bar{b}} \ &=  \ D_a\xi_{\bar{b}}-D_{\bar{b}}\xi_{a} -\xi^0 L_{a\bar{b}} \,, \\ 
\delta^{(0)} \phi \ &= \ \frac{1}{n}\partial_t \xi^0 \,, \\ 
\delta^{(0)} \varphi \ &=  \ \xi^0 \frac{\dot{\Phi}}{n} - \frac{1}{2} D_a\xi^a-\frac{1}{2}D_{\bar{a}}\xi^{\bar{a}} \,, \\ 
\delta^{(0)} \mathcal{A}^a \ &= \   \nabla_t \xi^a+L_{\bar{b}}{}^a \xi^{\bar{b}}  + D^a( \tilde{\xi}_0  + \xi^0) \,, \\ 
\delta^{(0)} \mathcal{A}^{\bar{a}} \ &= \ \nabla_t \xi^{\bar{a}} + L_b{}^{\bar{a}} \xi^b + D^{\bar{a}}(\tilde{\xi}_0 -  \xi^0) \,, 
\end{split} 
\end{align}  
where we performed the following rescaling of gauge parameters, 
\begin{align}
\begin{split}
\xi^0 \ \rightarrow \ \frac{1}{n} \xi^0 \,, \qquad 
\tilde{\xi}_0  \ \rightarrow \ n \tilde{\xi}_0 \,. 
\end{split}
\end{align} 
It is also  convenient to note that with respect to  spatial generalized diffeomorphisms with parameters $\xi^M=\bar{E}_{A}{}^{M}\xi^A$ the gauge 
transformations for $\varphi$ and ${\cal A}^M=\bar{E}_{A}{}^{M} {\cal A}^{A}$ simplify as follows: 
 \be
 \begin{split}
  \delta^{(0)}\varphi \ &= \  - \frac{1}{2} \partial_M \xi^M\;, \\
  \delta^{(0)}{\cal A}^{M} \ &= \ \frac{1}{n}\partial_t{\xi}^M\;.  
 \end{split}
 \ee
The gauge transformations in (\ref{quadrgauge1234}) act trivially for the special case:
\begin{equation}
\label{specialgauge}
\xi^a \ = \ D^a \chi \,, \quad \xi^{\bar{a}} \ = \ D^{\bar{a}} \chi \,, \quad \tilde{\xi}_0 = - \nabla_t \chi \,, 
\end{equation}
where $\chi$ is an arbitrary function. For $\delta^{(0)} h_{a\bar{b}}$ this can be seen by inspection, for $\delta^{(0)} \varphi$ by the constraint (\ref{weakconstraint}), and for $\delta^{(0)} \mathcal{A}^a$ and $\delta^{(0)} \mathcal{A}^{\bar{a}}$ by using the commutators  (\ref{Commutator a}) and (\ref{Commutator bar}).

\subsection{Canonical formulation of quadratic theory}
In order to elucidate the gauge structure of the quadratic double field theory on a time-dependent backgrounds 
we find it convenient to introduce a 
canonical formulation. We begin by computing the canonical momenta of the fields $\varphi$ and $h_{a\bar{b}}$:
\begin{equation}
\begin{split}
p_{\varphi} &\equiv \frac{\delta \mathcal{L}}{\delta \partial_t \varphi}  =  e^{-2\Phi}\big( - 8 \Gamma + 8 n^{-1}\dot{\Phi} \phi + 4 L^{a\bar{b}} h_{a\bar{b}} \big)\,,\\
p^{a\bar{b}}  & \equiv  \frac{\delta \mathcal{L}}{\delta \partial_t h_{a\bar{b}}} = e^{-2\Phi}\big( -2 \phi L^{a\bar{b}} - 2 \omega^{a\bar{b}} \big)\,.  
\end{split}
\end{equation}
In the following we rescale the dot by a factor of $n^{-1}$: $\dot{f} \equiv n^{-1} \partial_tf$. It is also convenient to rescale the canonical momenta by multiplying by the background quantity $e^{2\Phi}$:
\begin{equation}
\begin{split}
P_\varphi &\equiv e^{2\Phi} p_\varphi \ = \ - 8 \Gamma + 8  \dot{\Phi} \phi + 4 L^{a\bar{b}} h_{a\bar{b}} \;, \\
P_{a\bar{b}} &\equiv e^{2\Phi} p^{a\bar{b}} =  - 2 \phi L_{a\bar{b}} - 2 \omega_{a\bar{b}}  \,. 
\end{split} 
\end{equation}
Under the gauge transformations in (\ref{quadrgauge1234}), the canonical momenta transform as:
\begin{equation}
\begin{split}
\label{canmomGTs} 
\delta P_\varphi & = 4 \Delta \xi^0 + 4 L^{a\bar{b}} ( D_a \xi_{\bar{b}} - D_{\bar{b}} \xi_a ) \,, \\
\delta P_{a\bar{b}} & =4 \dot{\Phi} \xi^0 L_{a\bar{b}} - 4 D_a D_{\bar{b}} \xi^0  -2 L_{a}{}^{\bar{c}} (D_{\bar{c}} \xi_{\bar{b}} - D_{\bar{b}} \xi_{\bar{c}} ) + 2 L_{\bar{b}}{}^c (D_c \xi_a - D_a \xi_c) \,. 
\end{split}
\end{equation}
The Hamiltonian density is the Legendre transform of $\mathcal{L}$, defined as follows:
\begin{equation}
\mathcal{H} =   n\, e^{-2\Phi} ( P_\varphi   \dot{\varphi} + P^{a\bar{b}} \dot{h}_{a\bar{b}} )  - \mathcal{L}  \,, 
\end{equation}
which yields explicitly for the above fields 
\begin{equation}
\begin{split}
\mathcal{H} =   n\, e^{-2\Phi} \bigg\{ &- \frac{1}{4} P_{a\bar{b}} P^{a\bar{b}} - \frac{1}{16} P_{\varphi}^2 - \phi L_{a\bar{b}} P^{a\bar{b}} 
+ P^{a\bar{b}} (K_{a\bar{b}} + L_{a}{}^{c}h_{c\bar{b}} + L_{\bar{b}}{}^{\bar{c}} h_{a\bar{c}} )\\& - \frac{1}{2}\partial_M \mathcal{A}^M P_{\varphi}  +  \dot{\Phi} \phi P_{\varphi} + \frac{1}{2} ( L^{a\bar{b}} h_{a\bar{b}} ) P_{\varphi}
\\&
- 4  \dot{\Phi} ( L^{a\bar{b}} h_{a\bar{b}} ) \phi
 + L_{b}{}^{\bar{d}} L_{\bar{c}}{}^b h_{a}{}^{\bar{c}} h^{a}{}_{\bar{d}} +  L_{c}{}^{\bar{b}}L_{\bar{b}}{}^d h^{c}{}_{\bar{a}} h_{d}{}^{\bar{a}} + 2 L_{a}{}^{\bar{b}} L_{c}{}^{\bar{d}} h^{a}{}_{\bar{d}} h^{c}{}_{\bar{b}} \\
& - 2 L_{c}{}^{\bar{b}} h^{c\bar{a}} K_{\bar{a}\bar{b}} -2 L^{b\bar{c}} h^{a}{}_{\bar{c}} K_{ab} \\&
- 4 L^{a\bar{b}} K_{a\bar{b}} \varphi - ( L^{a\bar{b}} h_{a\bar{b}} ) ^2 - \mathcal{V}^{(2)}  \bigg\} \,.
\end{split}
\end{equation} 
In terms of this Hamiltonian density the complete quadratic action 
can be written with  the original fields and the canonical momenta in the following first-order form: 
\begin{equation}
S = \int dt \int d^{2d} {\bf X} \Big[n\, e^{-2\Phi} \big(P^{a\bar{b}}  \dot{h}_{a\bar{b}} + P_\varphi \dot{\varphi}\big)  - \mathcal{H} \Big]\,. 
\end{equation}
Naturally, upon integrating out the canonical momenta one recovers the original second-order action. 
Specifically, the equations of motion $E=0$ following from this action are encoded in the following components: 
\begin{flalign}\label{DefEternsors}
E_{P_\varphi} \ & = \ \frac{1}{8} P_\varphi +  \dot{ \varphi} + \frac{1}{2} \partial_M \mathcal{A}^M - \dot{\Phi} \phi - \frac{1}{2} L^{a\bar{b}} h_{a\bar{b}} \,,  \\ 
E_{Pa\bar{b}} \ & = \ \frac{1}{2} P_{a\bar{b}} + \nabla_t h_{a\bar{b}} - K_{a\bar{b}}  + \phi L_{a\bar{b}}   \,, \\  
E_\varphi \ &=  \ - \nabla_t P_\varphi + 2 \dot{\Phi} P_\varphi + 4 L^{a\bar{b}} K_{a\bar{b}} + 4 \Delta \phi - 8 \Delta \varphi +8 D^a D^{\bar{b}} h_{a\bar{b}} \,, \\ 
E_{a\bar{b}} \ &= \ - \nabla_t P_{a\bar{b}} + 2 \dot{\Phi} P_{a\bar{b}}   
 - \frac{1}{2} L_{a\bar{b}} P_\varphi  + 2 (L^{c\bar{d}} h_{c\bar{d}} ) L_{a\bar{b}} + 4 \dot{\Phi} \phi L_{a\bar{b}}\nonumber \\
 & \qquad 
-2 L_{\bar{d}}{}^c L_{c\bar{b}} h_{a}{}^{\bar{d}} - 2 L_{a}{}^{\bar{d}} L_{\bar{d}}{}^{c} h_{c\bar{b}} + 4 L_{a}{}^{\bar{d}} L_{\bar{b}}{}^{c} h_{c \bar{d}} 
+ 2 L_{a}{}^{\bar{d}} K_{\bar{b}\bar{d}} -2  L_{\bar{b}}{}^{c} K_{ac} \nonumber\\
&\qquad 
 + 8 D_aD_{\bar{b}} \varphi - 4 D_a D_{\bar{b}} \phi - 2\Delta h_{a\bar{b}} - 4D_a D^c h_{c\bar{b}} + 4 D_{\bar{b}} D^{\bar{c}} h_{a\bar{c}} \,, \\ 
E_a \ &= \ - D^{\bar{b}} P_{a\bar{b}} - \frac{1}{2} D_a P_\varphi  +4 L_{a\bar{b}} D^{\bar{b}} \varphi  + 2 L_{b}{}^{\bar{c}} D^b h_{a\bar{c}} - 2 L_{a}{}^{\bar{c}} D_b h^{b}{}_{\bar{c}}  \,, \\  
E_{\bar{a}} \ &= \ D^b P_{b\bar{a}} - \frac{1}{2} D_{\bar{a}} P_\varphi -4 L_{b\bar{a}} D^b \varphi -2 L_{c \bar{a}} D_{\bar{b}} h^{c \bar{b}} + 2 L_{c}{}^{\bar{b}} D_{\bar{b}} h^{c}{}_{\bar{a}}\,, \\ \label{DefEternsorsend}
E_\phi \ &= \ L_{a\bar{b}} P^{a\bar{b}} - \dot{\Phi} P_\varphi + 4 \dot{\Phi} L^{a\bar{b}} h_{a\bar{b}} + 4 \Delta \varphi - 4 D^a D^{\bar{b}} h_{a\bar{b}} \,. 
 \end{flalign} 
If one solves $E_{P_\varphi}=0$ and $E_{Pa\bar{b}}=0$ for the canonical momenta and substitutes the expressions into the action, one recovers the original Lagrangian (\ref{SimpleActionQuadLagrangian}). The tensors defining the equations of motion above satisfy 
Bianchi identities $G=0$ with the following components:  
\begin{flalign}
\label{Gh0}
&G^0 \equiv D_a E^a + D_{\bar{a}} E^{\bar{a}}  \,, \\
\label{Gsub0}
& G_0 \equiv D^{\bar{a}}E_{\bar{a}}  - D^a E_a - \dot{E}_\phi +2 \dot{\Phi} E_\phi + \dot{\Phi} E_\varphi  - L^{a\bar{b}} E_{a\bar{b}} + 4 \dot{\Phi} L^{a\bar{b}} E_{Pa\bar{b}} + 4 \Delta E_{P\varphi} - 4 D^a D^{\bar{b}} E_{Pa\bar{b}}   \,, \\
\label{Ga}
&G_a \equiv D^{\bar{b}} E_{a\bar{b}} - \nabla_t E_a + 2\dot{\Phi} E_a + L_{a}{}^{\bar{b}} E_{\bar{b}} + \frac{1}{2} D_a E_\varphi + 2L_{c}{}^{\bar{b}} D^c E_{Pa\bar{b}} - 2L_{a}{}^{\bar{b}} D^c E_{P c\bar{b}} + 4 L_{a}{}^{\bar{b}} D_{\bar{b}} E_{P \varphi}   \,, \\
\label{Gabar}
& G_{\bar{a}} \equiv - D^b E_{b\bar{a}}  - \nabla_t E_{\bar{a}} + 2\dot{\Phi}E_{\bar{a}} + L_{\bar{a}}{}^{b}E_b + \frac{1}{2} D_{\bar{a}} E_{\varphi}- 2 L_{\bar{b}}{}^c D^{\bar{b}} E_{Pc\bar{a}}  + 2 L_{\bar{a}}{}^c D^{\bar{b}} E_{Pc\bar{b }} + 4 L_{\bar{a}}{}^b D_bE_{P\varphi} . 
\end{flalign} 
For the explicit expressions for the $E$ tensors in (\ref{DefEternsors})--(\ref{DefEternsorsend}) the $G$ tensors vanish identically. 
This fact  expresses  the Bianchi identities. 
Furthermore, the Bianchi identities provide a straightforward proof of gauge invariance of the action, since 
one may verify with (\ref{quadrgauge1234}) and (\ref{canmomGTs}) that 
the gauge variation of the action can be written as the sum over the gauge parameter times the corresponding Bianchi identity:   
\begin{equation}
\delta^{(0)} S = \int dt \int d^{2d}{\bf X} \, n\,e^{-2\Phi} \big( \tilde{\xi}_0 G^0 + \xi^0 G_0 + \xi^a G_a + \xi^{\bar{a}} G_{\bar{a}} \big) = 0 \,.  
\end{equation} 

\subsection{Cubic theory}

We now turn to the cubic terms in the action obtained by expanding about a generic time-dependent background 
to third order in fluctuations. The corresponding Lagrangian  reads 
\begin{equation}
\begin{split}
\label{cubic action}
{\cal L}^{(3)}   =  n \, 
e^{-2\Phi} \bigg\{ & 2 \omega^{a\bar{b}} \omega_a{}^{c} h_{c\bar{b}} +2 \omega^{a\bar{b}} \omega_{\bar{b}}{}^{\bar{c}}h_{a\bar{c}}  - 2 L^{a\bar{b}}\omega^{c\bar{d}}  h_{a\bar{d}} h_{c\bar{b}} - 2 L^{a\bar{b}} K^{c\bar{d}} h_{c\bar{b}} h_{a\bar{d}} 
   + 4 L_{a\bar{b}}\omega^{a\bar{b}}\varphi^2  \\&
    + 2 \omega^{a\bar{b}} \mathcal{A}^M \partial_M  h_{a\bar{b}} + 8 ( \Gamma - \dot{\Phi} \phi  ) \mathcal{A}^M \partial_M \varphi  + 4L^{a\bar{b}} \varphi_{+}  \mathcal{A}^M \partial_M h_{a\bar{b}}   \\
 &-2 \varphi_{+}\mathcal{W}(\phi, \mathcal{A}, \varphi, h )  
 + \mathcal{V}^{(3)}(\phi, \varphi_{-}, h) \bigg\} \,, \\
\end{split}
\end{equation}
where we have grouped objects familiar from the quadratic action into
\begin{align}
\begin{split}
\mathcal{W}(\phi, \mathcal{A}, \varphi, h )   \ = \ & 
 -4 \Gamma^2
  - \omega_{a\bar{b}}\omega^{a\bar{b}} 
  + \frac{1}{2}\big( \omega_{ab}\omega^{ab} + \omega_{\bar{a}\bar{b}} \omega^{\bar{a}\bar{b}}  \big) - \frac{1}{2}\big( K_{ab}K^{ab} + K_{\bar{a}\bar{b}} K^{\bar{a}\bar{b}}  \big)\\
 & +2 \phi (4  \dot{\Phi} \Gamma- L_{a\bar{b}}  \omega^{a\bar{b}}   )     \,. 
\end{split}
\end{align}
Moreover, we defined 
\begin{equation}
 \mathcal{V}^{(3)} ( \phi, \varphi_{-}, h)= \mathcal{T}+ \mathcal{U} \,,
\end{equation}
where
\begin{align}
\begin{split}
\mathcal{T} \  = \ &  4 h^{a\bar{b}} ( D_a h^{c\bar{d}} D_{\bar{b}}h_{c\bar{d}} - D_a h^{c\bar{d}} D_{\bar{d}} h_{c\bar{b}} - D_{\bar{b}} h^{c\bar{d}} D_{c} h_{a\bar{d}} ) \\
 &+4\varphi_{-}(  D^a h^{c\bar{d}} D_a h_{c\bar{d}} + D^a h_{a\bar{b}} D_c h^{c\bar{b}} - D^{\bar{b}} h_{a\bar{b}} D_{\bar{c}} h^{a\bar{c}} + 2 h_{a\bar{b}} D^a D_c h^{c\bar{b}} - 2 h_{a\bar{b}} D^{\bar{b}} D_{\bar{c}} h^{a\bar{c}} ) \\
 & - 16h_{a\bar{b}} \varphi_{-} D^{a}D^{\bar{b}} \varphi_{-} - 8 \varphi_{-}^2 D^a D_a \varphi_{-} \,, \\
 \mathcal{U} \ = \ & 4 \varphi_{-}D^a\phi D_a\phi + 8 \varphi_{-}\phi D^a D_a \phi - 4 \phi D^a\phi D_a \phi + 4 h^{a\bar{b}} \phi D_a D_{\bar{b}} \phi \,. \\ 
\end{split}\end{align}
We note that $\mathcal{T}$ matches the form of the cubic expansion of the double field theory action around a constant background without space-time split in \cite{Hohm:2011dz}, with $\varphi_{-}$ playing the role of the dilaton. 

Finally, expanding the full gauge transformations (\ref{gauge_transfr_reparametrized}) to first order in fields (i.e.~including 
all terms quadratic in fields and gauge parameters) one finds: 
\begin{align}\label{nonlinearGaugeTrans}
\begin{split}
\delta^{(1)} h_{a\bar{b}} & \ = \ \xi^0 \omega_{a\bar{b}} + \xi^N \partial_N h_{a\bar{b}} + ( D_{\bar{b}} \xi^{\bar{c}} - D^{\bar{c}} \xi_{\bar{b}} ) h_{a\bar{c}} + ( D_a \xi^c - D^c \xi_a ) h_{c\bar{b}}  \,, \\
\delta^{(1)}  \phi & \ = \ \frac{1}{n} \partial_t (  \xi^0 \phi ) - \mathcal{A}^M \partial_M \xi^0 + \xi^M \partial_M \phi \,, \\
\delta^{(1)} \varphi & \ = \ \xi^0 \Gamma + \xi^N \partial_N  \varphi  \,,\\
\delta^{(1)}  \mathcal{A}^a & \ = \ - 2 h^{a\bar{b}}  D_{\bar{b}} \xi^0 +2 \phi D^a \xi^0  + \xi^B D_B \mathcal{A}^a +  (D^a\xi_{B} - D_B \xi^a)\mathcal{A}^B \,, \\ 
\delta^{(1)}  \mathcal{A}^{\bar{a}} & \ = \ - 2  h^{b\bar{a}}  D_b \xi^0  -2 \phi  D^{\bar{a}} \xi^0  + \xi^B D_B \mathcal{A}^{\bar{a}} +  (D^{\bar{a}}\xi_{B} - D_B \xi^{\bar{a}})\mathcal{A}^B  \,.
\end{split} 
\end{align}
Gauge invariance of the action to cubic order in fluctuations requires 
\begin{equation}
 \delta^{(1)} S^{(2)} + \delta^{(0)}  S^{(3)} = 0  \,, 
\end{equation} 
as the reader may  verify  with the above formulas by a straightforward but tedious computation.

\section{Dualities and cocycles}

In this section we discuss the duality properties of double field theory in time-dependent backgrounds 
and the role of cocycle factors in the weakly constrained theory. After discussing in the first subsection 
some general features of duality transformations of time-dependent backgrounds, in the second subsection 
we turn to a special class of Friedmann-Robertson-Walker (FRW) backgrounds 
and discuss their invariance group. In the final subsection 
we point out that the theory to cubic order is well-defined in a 
weakly constrained sense, for which string theory suggests the presence of cocycle factors, which we discuss.

\subsection{Dualities in time-dependent backgrounds }

Double field theory is manifestly $O(d,d)$ invariant and so are the quadratic and cubic actions 
about time-dependent backgrounds that we determined above. More precisely, this $O(d,d)$ invariance 
implies, in general, the following \textit{duality property}: changing the background $E_{A}{}^{M}(t)$
by an $O(d,d)$ transformation leads a physically equivalent theory. This follows immediately from the 
$O(d,d)$ invariance of the action since the $O(d,d)$ transformation of $E_{A}{}^{M}(t)$ can then be absorbed 
by an inverse transformation of the fields, which in turn can be viewed as a field redefinition. 
Thus, the actions for two backgrounds related by an $O(d,d)$ transformation are equal up to field redefinitions 
and hence physically equivalent. 

While this result is guaranteed by the general double field theory formalism, it is instructive 
to give a parametrization of the most general time-dependent background and to inspect 
the duality properties explicitly. 
A general time-dependent metric $G_{ij}(t)$ and Kalb-Ramond B-field can be packaged into 
 \begin{equation}\label{calE}
  {\cal E}_{ij}(t)  =  G_{ij}(t)  + B_{ij}(t)\;. 
 \end{equation}
In order to write a general frame $E_{A}{}^{M}(t)$ \textit{without gauge fixing} one needs to introduce 
on top of this two independent  $d$-dimensional frame fields $e_{a}{}^{i}(t)$ and $\bar{e}_{\bar{a}}{}^{i}(t)$, 
for then we can write 
 \begin{equation}\label{GeneralFRAMEE}
{E}_A{}^M = 
\begin{pmatrix} {E}_{ai} & {E}_a{}^i \\ {E}_{\bar{a} i } & { E}_{\bar{a}}{}^i \end{pmatrix} 
= \frac{1}{\sqrt{2}} \begin{pmatrix} {\cal E}_{ji} e_{a}{}^{j}  & - e_a{}^i \\ {\cal E}_{ij} 
\bar{e}_{\bar{a}}{}^{j} &\bar{e}_{\bar{a}}{}^i \end{pmatrix} \;, 
\end{equation}
or, in matrix notation with $E=(E_{A}{}^{M})$, $e=(e_{a}{}^{i})$ and $\bar{e}=(\bar{e}_{\bar{a}}{}^{i})$, 
 \be\label{EmatrixComponents}
  E = \frac{1}{\sqrt{2}} \begin{pmatrix} e{\cal E} & - e \\ \bar{e}{\cal E}^t 
 &\bar{e} \end{pmatrix} \;. 
 \ee
This background satisfies the constraint (\ref{flatetaisG}) in that  
 \begin{equation}\label{flatframemetric}
  {\mathcal{G}}_{AB} = \begin{pmatrix} -e_{a}{}^{i} e_{b}{}^{j} G_{ij} & 0 \\ 
  0 & \bar{e}_{\bar{a}}{}^{i} \bar{e}_{\bar{b}}{}^{j} G_{ij} \end{pmatrix} \;. 
 \end{equation}
This parametrization of the frame preserves  the full $O(d,d)$ and $GL(d)\times GL(d)$ covariance, 
since we have $3d^2$ degrees of freedom
($e$, $\bar{e}$ and ${\cal E}$), as it should be for a frame with $(2d)^2=4d^2$ components satisfying the $d^2$ constraints 
(\ref{flatetaisG}) (i.e., the $GL(d)\times GL(d)$ covariant constraints that the off-diagonal blocks of (\ref{flatframemetric}) vanish).

Under global $O(d,d)$  and local (time-dependent) $GL(d)\times GL(d)$ transformations 
this frame transforms as 
 \begin{equation}
  E'_{A}{}^{M}(t)  =  
  \Lambda_{A}{}^{B}(t) E_{B}{}^{N}(t) h^{M}{}_{N}\;, \qquad h = \begin{pmatrix}  a & b \\ c & d \end{pmatrix} 
  , \quad \Lambda_A{}^{B} =  \begin{pmatrix}  \Lambda_a{}^b & 0 \\ 0 & \bar{\Lambda}_{\bar{a}}{}^{\bar{b}} \end{pmatrix}\;, 
 \end{equation}
or, in matrix notation, as $E'=\Lambda E h^t$.  Note that the transformation does not act on any coordinate argument 
since the background depends only on time on which $O(d,d)$ does not act.
In the matrix notation (\ref{EmatrixComponents}) this yields  
 \begin{equation}\label{Mmatrices}
  e'  =  \Lambda \, e\, M \;, \qquad \bar{e}'  =  \bar{\Lambda} \, \bar{e}\, \bar{M}\;,
  \qquad {\cal E}' = (a{\cal E}+b)(c{\cal E}+d)^{-1}\,, 
 \end{equation}
where 
 \be
   M(t)  :=  d^t- {\cal E}(t) c^t\;, \qquad \bar{M}(t) := d^t+{\cal E}(t)^t c^t\;. 
 \ee
The matrices $M$ and $\bar{M}$ are familiar from the double field theory on flat space \cite{Hull:2009mi}, 
with the important difference that here they are time-dependent. 
We could now completely fix the background $GL(d)\times GL(d)$ frame transformations by setting $e=\bar{e}={\bf 1}$, 
which in turn would require compensating frame transformations given by $\Lambda=M^{-1}$ and $\bar{\Lambda}=\bar{M}^{-1}$, 
as follows from (\ref{Mmatrices}). However, due to (\ref{flatframemetric}) this gauge choice leads to a time-dependent 
tangent space metric, so that the operations of raising and lowering indices do not commute with time derivatives. 
It is instead more convenient to choose a gauge for which the tangent space metric is constant, as done in the 
previous sections. We will next consider particular frames with this property.

\subsection{Friedmann-Robertson-Walker backgrounds }\label{FRWbackgrounds}

We now turn to the class of time-dependent backgrounds with the largest degree of symmetry: 
the  Friedmann-Robertson-Walker (FRW) spaces with vanishing spatial curvature. 
These are characterized by a single time-dependent function, 
the scale factor $a(t)$, while the B-field vanishes. Specifically, (\ref{calE}) then reduces to ${\cal E}_{ij}(t)=a^2(t)\delta_{ij}$. 
It will furthermore  be convenient to introduce two constant but otherwise arbitrary bases or frames $e_{a}{}^{i}$ and 
$\bar{e}_{\bar{a}}{}^i$ for the doubled spatial geometry. Since these bases are unconstrained, the `tangent space' 
metrics defined in terms of the flat spatial metric by 
 \be\label{flatMetrics}
  g_{ab} = e_{a}{}^{i} e_{b}{}^{j} \delta_{ij}\,, \qquad 
  g_{\bar{a}\bar{b}} = \bar{e}_{\bar{a}}{}^{i} \bar{e}_{\bar{b}}{}^{j} \delta_{ij}\,, 
 \ee
are then  independent constant metrics (of Euclidean signature). As usual, we use $\delta_{ij}$ and $\delta^{ij}$ to 
lower and raise indices $i,j,\ldots$, while $g_{ab}$ and $g^{ab}$, respectively $g_{\bar{a}\bar{b}}$ and $g^{\bar{a}\bar{b}}$, 
are used to lower and raise indices $a,b,\ldots$ and $\bar{a},\bar{b},\ldots$.
Using the frames $e_{a}{}^{i}$ and 
$\bar{e}_{\bar{a}}{}^i$, together with their inverses denoted by 
 $e_{i}{}^{a}$ and $\bar{e}_{i}{}^{\bar{a}}$, to convert indices $a,b,\ldots$ and $\bar{a},\bar{b},\ldots$ to $i,j,\ldots$, 
one may verify that the different operations of raising and lowering indices are mutually compatible. For instance, in 
 \be
  e^{ia} = \delta^{ij} e_{j}{}^{a} = g^{ab} e_{b}{}^{i}\;, 
 \ee
the second equation follows by contraction with $e_{a}{}^{k}$ and using  the inverse of (\ref{flatMetrics}). 
Notably, there are tensors 
 \be\label{mixedtensorsg}
  g_{a}{}^{\bar{b}} := e_{a}{}^{i} \,\bar{e}_{i}{}^{\bar{b}}\,, \qquad 
  g_{\bar{a}}{}^{b} := \bar{e}_{\bar{a}}{}^{i}\, e_{i}{}^{b}\, 
 \ee
that connect unbarred and barred indices and that satisfy 
\begin{equation}
\begin{split} 
g_{a}{}^{\bar{b}} \,g_{\bar{b}}{}^c = \delta_a{}^c \,, \quad {\rm etc.} 
\end{split}
\end{equation}

With these objects we can write the full background frame as 
\begin{equation}\label{FRWdoubledFrame}
{E}_A{}^M(t) = \frac{1}{\sqrt{2}} \begin{pmatrix} a(t) \delta_{ij}e_{a}{}^{j} & -a^{-1}(t) e_a{}^i \\ a(t)\delta_{ij} \bar{e}_{\bar{a}}{}^{ j } & a^{-1}(t) \bar{e}_{\bar{a}}{}^i \end{pmatrix}  \,. 
\end{equation}
This is of the general form (\ref{GeneralFRAMEE}), except that the frames $e_{a}{}^{i}$ and 
$\bar{e}_{\bar{a}}{}^i$ have been rescaled by $a(t)$ to make them constant. 
The corresponding tangent space metric then satisfies the constraint  (\ref{flatetaisG}) with the constant 
metric on the doubled tangent space
\begin{equation}\label{FRWtangentspace} 
 \mathcal{G}_{AB} = 
\begin{pmatrix}
- g_{ab} & 0 \\
0 & g_{\bar{a}\bar{b}} 
\end{pmatrix}\,. 
\end{equation}
Note the relative sign in the upper-left block, which will lead to  a change of convention in raising 
and lowering indices in sec.~4 below. 
The differential operators $D_A=(D_a, D_{\bar{a}})\equiv E_{A}{}^{M}\partial_M$ are given by 
 \be
  \begin{split}
    D_{a} &= -\tfrac{1}{\sqrt{2}}\left(a^{-1}(t) \partial_a  -   a(t)\tilde{\partial}_a\right)\;, \\
    D_{\bar{a}} &= \tfrac{1}{\sqrt{2}}\left(a^{-1}(t)\partial_{\bar{a}}\ + \ a(t) \tilde{\partial}_{\bar{a}}\right)\;, 
  \end{split}
 \ee
using the notation $\partial_a=e_{a}{}^{i}\partial_{i}$, $\tilde{\partial}_a=e_{a}{}^{i}\tilde{\partial}_i$, 
and similarly for barred indices. Note that despite the notation there is only one kind 
of momentum and one kind of winding derivative: with (\ref{mixedtensorsg}) we have 
$\partial_{\bar{a}}=g_{\bar{a}}{}^{b}\partial_b$ and $\tilde{\partial}_{\bar{a}}=g_{\bar{a}}{}^{b}\tilde{\partial}_b$.

We can next give the explicit form of the tensor  
\begin{equation}
L_{A}{}^{B} \equiv  \dot{E}_{A}{}^M E_{M}{}^{B } \,, 
\end{equation}
used in the main text, where we recall the notation $\dot{X} \equiv n^{-1} \partial_t X$. Using 
(\ref{FRWdoubledFrame}) one finds for its components 
\begin{equation}\label{Ltensorcomp}
\begin{split}
L_{a}{}^{\bar{b}} & = H  g_{a}{}^{\bar{b}}  \,,\qquad 
L_{\bar{a}}{}^{b} = Hg_{\bar{a}}{}^{b} \,,\qquad 
L_{a}{}^{b}   = L_{\bar{a}}{}^{\bar{b}} = 0 \,,
\end{split}
\end{equation}
with Hubble parameter 
 \be
   H \equiv   \frac{\dot{a}}{a}\;. 
 \ee  
With this choice of backgrounds, the background equations of motion in (\ref{eomLdot})-(\ref{eomdotPhi LL}) become:
\begin{flalign}
\label{simpbkgrndeom1}
(\dot{H}-2\dot{\Phi} H)g_{a}{}^{\bar{b}}  =0\,,\\
\label{simpbkgrndeom2}
-4 \ddot{\Phi} + 4 \dot{\Phi}^2 + d H^2 = 0 \,,\\
\label{simpbkgrndeom3}
4\dot{\Phi}^2 - d H^2 = 0 \,. 
\end{flalign} 
As done before in sec.~2, we can add (\ref{simpbkgrndeom2}) and (\ref{simpbkgrndeom3}) to obtain the equation
\begin{equation}
- \ddot{\Phi} + 2 \dot{\Phi}^2 = 0 \,. 
\end{equation}
The equations of motion imply that the following quantity is conserved,
\begin{equation}
\beta  \equiv  \frac{\dot{\Phi}}{H} \,, 
\end{equation}
since by taking its time derivative we have
\begin{equation}
\dot{\beta} = \frac{(\ddot{\Phi}H -  \dot{\Phi}\dot{H})}{H^2} = \frac{2\dot{\Phi}^2 H - \dot{\Phi}(2\dot{\Phi}H)}{H^2} =0 \,. 
\end{equation}

Let us next analyze the invariance groups of these FRW  backgrounds. We 
first note that since we are now considering a rather special class of backgrounds, under a 
general $O(d,d)$ or $GL(d)\times GL(d)$ 
transformation the background frame will of course not stay in the same class. 
However, under constant or time independent $GL(d)\times GL(d)$  transformations 
the above backgrounds transform into themselves, just with the frames $e_{a}{}^{i}$ and 
$\bar{e}_{\bar{a}}{}^i$ rotated, which follows  as in (\ref{Mmatrices}). 
The genuine duality transformation left in $O(d,d)$ is given by 
 \begin{equation}
  h = \begin{pmatrix}  a & b \\ c & d \end{pmatrix} = \begin{pmatrix}  0 & 1 \\ 1 & 0 \end{pmatrix} : \qquad
  {\cal E}'(t) = {\cal E}^{-1}(t) \qquad \Leftrightarrow \qquad  a'(t) = \frac{1}{a(t)}\;, 
 \end{equation}
as follows with the last relation in (\ref{Mmatrices}).\footnote{Depending on the form of $e$ and $\bar{e}$ 
this transformation requires compensating $GL(d)\times GL(d)$  transformations. For instance, if $e=\bar{e}={\bf 1}$ 
then the compensating transformations are $\Lambda=-{\cal E}^{-1}$, $\bar{\Lambda}=({\cal E}^t)^{-1}$, 
as can be verified  with (\ref{Mmatrices}).}
This is the expected T-duality or scale-factor duality property of string cosmology.

We can then turn to the problem of determining the \textit{invariance group}, i.e., the subgroup of $O(d,d)$ 
(possibly accompanied by  $GL(d)\times GL(d)$  transformations) for which $E'=E$. For flat space backgrounds, 
which are contained in the above class of backgrounds for $a\equiv 1$, the invariance group is $O(d)\times O(d)$. 
Thus, the invariance group of genuine  FRW backgrounds must be a subgroup of $O(d)\times O(d)$, which in turn 
is embedded into $O(d,d)$ as 
 \begin{equation}
  h \ = \ \frac{1}{2}\begin{pmatrix} \ell+\bar{\ell} &  \bar{\ell}-\ell\\  \bar{\ell}-\ell & \ell+\bar{\ell}  \end{pmatrix} , 
  \qquad 
   \ell\in O(d)\;, \quad \bar{\ell}\in O(d) \,. 
 \end{equation}
 Therefore, for ${\cal E}_{ij}=a^2(t)\delta_{ij}$ one finds for the factors in the transformation formula (\ref{Mmatrices})
  \be
   a{\cal E}+b = \frac{1}{2}\left( (\ell+\bar{\ell} )a^2(t) + \bar{\ell} -\ell \right)\,,
   \qquad
   c{\cal E}+d = \frac{1}{2}\left((\bar{\ell}-\ell)a^2(t) +\ell+\bar{\ell}\right)\;. 
  \ee
For flat backgrounds with $a\equiv 1$ this yields $a{\cal E}+b=\bar{\ell}$ and $c{\cal E}+d=\bar{\ell}$ 
and hence ${\cal E}'= \bar{\ell}\bar{\ell}^{-1} = {\bf 1}={\cal E}$. This confirms that $O(d)\times O(d)$
is an invariance group of ${\cal E}={\bf 1}$.  
This argument fails if $a(t)$ is a genuine function of time, but in this case we note that upon identifying 
$\ell=\bar{\ell}$ we have $a{\cal E}+b=a^2(t) \ell$ and $c{\cal E}+d = \ell$, so that the background is invariant: 
 \be
   {\cal E}'(t)=a^2(t) \ell \ell^{-1} = a^2(t){\bf 1}= {\cal E}(t)\;.
 \ee  
Thus, the invariance group of generic FRW backgrounds is given by the \textit{diagonal} subgroup 
${\rm diag}(O(d)\times O(d))$, as mentioned in the introduction.

\subsection{Weak constraint and cocycles}

We now discuss in more detail the weak  constraint originating from the level-matching constraint 
of string theory on toroidal backgrounds and explain how the cubic theory is consistent as a weakly constrained theory. 
We denote a generic field by $\phi$ (that need not be the lapse fluctuation above) and expand 
in Fourier modes of the torus as 
 \be\label{FoueriSeries}
  \phi(t,{\bf X}) = \sum_{{\bf K}\in \mathbb{Z}^{2d}}\phi_{\bf K}(t) \,e^{i {\bf K}^t{\bf X}}\;. 
 \ee
We find it convenient to take the sum over ${\bf K}\in \mathbb{Z}^{2d}$ to be unconstrained
and to implement the weak constraint $\partial^{M}\partial_M\phi=0$ by demanding 
\be\label{modeconstraint}
 {\bf K}^2\equiv \eta^{MN}{\bf K}_M{\bf K}_{N} \neq 0 \quad \Rightarrow \quad \phi_{\bf K}=0\;. 
\ee
Put differently, we take the Fourier modes $\phi_{\bf K}$ to be non-zero only if the corresponding 
Fourier label ${\bf K}$ is null with respect to the $O(d,d)$ metric. 
The weak constraint is then obeyed for (\ref{FoueriSeries}).

The consistency problem for a weakly constrained theory arises because the usual point-wise 
product of  functions satisfying the weak constraint in general does not satisfy the weak constraint. 
In string field theory this consistency problem is resolved by having a modified   product, which 
simply projects out all Fourier modes that do not obey (\ref{modeconstraint}). 
Specifically, we define the product 
 \be\label{bulletproduct}
  \big(\phi^1\bullet  \phi^2\big)(t, {\bf X}) \equiv 
    \sum_{{\bf K}_1,{\bf K}_2\in \mathbb{Z}^{2d}}\delta_{{\bf K}_1\cdot {\bf K_2},0}\, 
  C({\bf K}_1,{\bf K}_2)\, 
  \phi_{{\bf K}_1}^1(t) \phi_{{\bf K}_2}^2(t) e^{i({\bf K}_1+{\bf K}_2)^t{\bf X}}\;, 
 \ee 
where a Kronecker delta was introduced to set to zero any term that does not obey ${\bf K}_1\cdot{\bf K}_2 =  0$. 
Since by (\ref{modeconstraint}) any term in the above sum vanishes unless ${\bf K}_1^2={\bf K}_2^2=0$
it then follows that only terms with $({\bf K}_1+{\bf K}_2)^2=0$ have non-vanishing coefficients. Therefore, 
the $\bullet$ product of two functions obeying the weak constraint also obeys the weak constraint: 
$\partial^M\partial_M(\phi^1\bullet  \phi^2\big)=0$.  
The so-called cocycle function $C({\bf K}_1,{\bf K}_2)$ appearing in the definition  is a sign factor that is symmetric in its two arguments
and defined in terms of the components of ${\bf K}=(w,k)$ by 
 \be\label{CocylceDEF}
  C({\bf K}_1,{\bf K}_2) \equiv (-1)^{k_1w_2} \equiv  (-1)^{{\bf K}_1^t Q{\bf K}_2}\;, \qquad   Q^{MN} \equiv  \begin{pmatrix}   0 & 0 \\
  1 & 0  \end{pmatrix}\;, 
 \ee 
where $kw$ denotes the usual (Euclidean) dot product between $d$-dimensional vectors.  
More precisely, given the cocycle's  role in the deformed product (\ref{bulletproduct}), 
it only needs to be defined for arguments that are already null and mutually orthogonal. 
We then have  $0={\bf K}_1\cdot{\bf K}_2=k_1w_2+w_1k_2$ and hence $(-1)^{k_1w_2}=(-1)^{k_2w_1}$, 
so that the cocycle is indeed symmetric. Similarly, if ${\bf K}_1$ is proportional to ${\bf K}_2$ we have 
$k_1 w_2\propto k_1w_1=0$ and hence 
 \be\label{cocyleIsone}
  {\bf K}_1\propto {\bf K}_2\quad \Rightarrow \quad C({\bf K}_1,{\bf K}_2) = 1\;. 
 \ee
Let us also record the formula for the mode of the product  with Fourier label ${\bf K}$: 
 \be\label{FourierModeProduct}
  (\phi^1\bullet \phi^2)_{\bf K} = \sum_{{\bf P}\in \mathbb{Z}^{2d}} \delta_{{\bf P}\cdot{\bf K},0}\, C({\bf P}, {\bf K}) \,
  \phi_{{\bf P}}^1\cdot \phi_{{\bf K}-{\bf P}}^2\;, 
 \ee
where we used $C({\bf P}, {\bf K}-{\bf P})=C({\bf P}, {\bf K})$, which  follows quickly with (\ref{CocylceDEF}). 
Since the tensor $Q^{MN}$ in (\ref{CocylceDEF}) is \textit{not} $O(d,d)$ invariant a theory employing the $\bullet$ product 
seems to violate the $O(d,d,\mathbb{Z})$ duality invariance, but we will see below that the duality is realized 
through  a non-standard action involving a similar sign factor.  

We now discuss some properties of the $\bullet$ product. Since the cocycle is symmetric, the product is still commutative. 
Furthermore, it is easy to convince oneself that the modified product (\ref{bulletproduct}) still obeys the Leibniz rule: 
 \be\label{bulletLeibniz}
  \partial_M(\phi^1\bullet \phi^2) = (\partial_M\phi^1)\bullet \phi^2 + \phi^1\bullet (\partial_M\phi^2)\;. 
 \ee
The product  is, however, not associative, as can be seen  by considering  the following three `pure Fourier mode' 
functions for $d=1$:\footnote{We thank Henning Samtleben for suggesting  this example.}  
 \be
  f = e^{i\tilde{x}}\;, \quad g=e^{ix}\;, \quad h=e^{-ix}\;, 
 \ee
which all, depending only on $x$ or only on $\tilde{x}$, 
obey the weak constraint. Moreover, since these functions thus carry only momentum 
or only winding, the cocycle factor between any two of them is trivially unity. Since the point-wise product $f\cdot g=e^{ix+i\tilde{x}}$ 
violates the weak constraint the $\bullet$ product is just zero, $f\bullet g=0$, and thus 
  $(f\bullet g)\bullet h = 0$.  
On the other hand, $g\cdot h=1$ obeys the weak constraint so that  $g\bullet h=1$ and hence  
  $f\bullet (g\bullet h) = f$.  
Thus, associativity does not hold.  While associativity does not hold 
in general it does hold when integrated over the whole (doubled) torus: 
 \be\label{integratedASS}
  \int_{T^{2d}} d^{2d}{\bf X}\Big[(\phi^1\bullet \phi^2)\bullet \phi^3-\phi^1\bullet (\phi^2\bullet \phi^3)\Big] = 0\;. 
 \ee
In order to see this we use $\frac{1}{(2\pi)^{2d}}\int d^{2d}{\bf X}\,e^{i{\bf K}\cdot {\bf X}}=\delta({\bf K})$, 
where $\delta({\bf K})$ is the Kronecker delta that is equal to $1$ if all components of ${\bf K}$ are zero 
and zero otherwise. 
This implies for the integral of a generic scalar with Fourier series (\ref{FoueriSeries}): 
 \be\label{zeromodeInt}
  \int dt \int d^{2d}{\bf X}\,\phi(t,{\bf X}) = (2\pi)^{2d} \int dt \,\phi_{{\bf 0}}(t)\;,
 \ee 
i.e.~the integral reduces to that over the zero mode with ${\bf K}={\bf 0}$. (In the following we will omit 
the time integrals as they play no role.) 
Using this we compute 
 \be\label{quadraticidentity}
  \int_{T^{2d}} d^{2d}{\bf X}\, (\phi^1\bullet \phi^2) = (2\pi)^{2d} \sum_{{\bf K}\in \mathbb{Z}^{2d}} \phi_{-{\bf K}}^1\, \phi_{{\bf K}}^2
  =  \int_{T^{2d}} d^{2d}{\bf X}\,(\phi^1\cdot \phi^2)\;. 
 \ee
Here we used that the Kronecker delta enforces $ {\bf K}_2=-{\bf K}_1$ in (\ref{bulletproduct}), 
so that ${\bf K}_1\cdot {\bf K}_2=0$ automatically, and the cocycle factor is $1$ due to (\ref{cocyleIsone}). 
Therefore, under an integral the modified product equals the ordinary product. 
Similarly, 
\be\label{AssSTEP}
\begin{split}
  \int_{T^{2d}} d^{2d}{\bf X}\Big[(\phi^1\bullet \phi^2)\bullet \phi^3\Big]
  &= \int_{T^{2d}} d^{2d}{\bf X}\Big[(\phi^1\bullet \phi^2)\cdot \phi^3\Big] 
  =(2\pi)^{2d} \sum_{{\bf K}\in \mathbb{Z}^{2d}}(\phi^1\bullet \phi^2)_{\bf K}\cdot \phi^3_{-{\bf K}}\\
  &= (2\pi)^{2d} \sum_{{\bf K}, {\bf P} \in \mathbb{Z}^{2d}}
   \delta_{{\bf P}\cdot{\bf K},0}\, C({\bf P}, {\bf K}) \,
  \phi_{{\bf P}}^1\cdot \phi_{{\bf K}-{\bf P}}^2\cdot \phi_{-{\bf K}}^3
  \;, 
 \end{split}
 \ee
where we used (\ref{FourierModeProduct}).  Recalling the convention (\ref{modeconstraint}) we can use 
${\bf P}^2=({\bf K}-{\bf P})^2={\bf K}^2=0$ and hence ${\bf P}\cdot {\bf K}=0$. 
It is then easy to verify that the second term in (\ref{integratedASS}), i.e., the integral over $\phi^1\bullet (\phi^2\bullet \phi^3)$, 
equals (\ref{AssSTEP}). This proves (\ref{integratedASS}).

We can now see that the cubic theory is consistent as a weakly constrained theory. 
For the free part this is  immediate: in the quadratic action, due to (\ref{quadraticidentity}), 
the regular point-wise product equals the modified product. 
Similarly, the \textit{linear} gauge transformations needed to check invariance 
of the quadratic action are consistent with the weak constraint since the gauge parameters also obey the weak constraint. 
(Note that the spatial integral  is always the ordinary integral over the full doubled torus; only the fields are assumed to 
be restricted by the weak constraint.)  
Turning to the cubic theory, we first note that due to (\ref{integratedASS}) there is no ambiguity in writing 
the terms of cubic order in the action with the $\bullet$ product. 
Furthermore, all quadratic terms in the non-linear gauge transformations (\ref{nonlinearGaugeTrans}) must be 
reinterpreted as employing the $\bullet$ product in the quadratic terms, but since to this order all terms in the gauge transformed 
action are still at most cubic in fields and parameters, gauge invariance
follows as before, using  (\ref{bulletLeibniz}) and (\ref{integratedASS}). 
It should be emphasized that this argument to cubic order goes through whether a cocycle factor is included  or not. 
Thus, as far as the cubic truncation is concerned there is no need to include a cocycle factor, but the claim in 
the string field theory literature is that to quartic and higher order gauge invariance requires the cocycle factors \cite{Hata:1986mz,Maeno:1989uc,Kugo:1992md}, 
although it remains a possibility  that for the pure DFT sector a theory 
without cocycles may exist.\footnote{OH thanks Barton Zwiebach for collaboration 
on the issues discussed here.}

We close this section by returning to the $O(d,d,\mathbb{Z})$ duality invariance, 
whose realization is non-standard in the presence of 
cocycle factors \cite{Kugo:1992md}.
To understand this we first consider 
the standard case without cocycles, for which $h\in O(d,d,\mathbb{Z})$ acts as 
 \be
  \phi(t,{\bf X}) \ \rightarrow \ \phi'(t,{\bf X}') = \rho(h) \phi(t,{\bf X})\,, \quad \text{where} \quad 
  {\bf X}' = h{\bf X}\;. 
 \ee
Here $\rho(h)$ denotes the matrix in the (finite-dimensional) $O(d,d)$ representation of $\phi$. Expanding both sides 
into Fourier modes according to (\ref{FoueriSeries}) then yields for the Fourier components 
 \be\label{NaiveTransformation}
  \phi_{{\bf K}'}'(t) = \rho(h) \phi_{\bf K}(t)\,, \quad \text{where} \quad 
  {\bf K}'=(h^t)^{-1} {\bf K}\,. 
 \ee
Note that this transformation rule is the reason that the duality group is the discrete $O(d,d,\mathbb{Z})$:   
Since the components of ${\bf K}$ are integers, the matrix components of $h\in O(d,d)$ must be integers too 
in order to preserve this property, hence breaking  $O(d,d,\mathbb{R})$ to $O(d,d,\mathbb{Z})$. 
 
Any conventional contractions of products of $O(d,d)$ tensors are covariant under this transformation, 
but since the modified product (\ref{FourierModeProduct}) involves the cocycle factor that is not $O(d,d)$ invariant, 
the modified product is not covariant under the standard transformation. 
We will now show, however, that the modified product  is covariant under a similarly modified $O(d,d,\mathbb{Z})$ action. 
Specifically, (\ref{NaiveTransformation}) is modified to 
  \be\label{ModTransformation}
  \phi_{{\bf K}'}'(t) = (-1)^{F({\bf K},h) } \rho(h) \phi_{\bf K}(t)\,,
 \ee
with a sign factor depending on ${\bf K}$ and $h\in O(d,d)$ to be determined. 
We begin by computing the anomalous transformation of the cocycle factor (\ref{CocylceDEF}) 
under 
\be 
  {\bf K}  \rightarrow  {\bf K}'  =  (h^t)^{-1}{\bf K}\;, \qquad h  =  \begin{pmatrix}   a & b \\
  c & d  \end{pmatrix}  \in  O(d,d,\mathbb{Z})\;, 
\ee
i.e., 
 \be\label{cocyctrans}
  C({\bf K}_1',{\bf K}_2') \ = \ 
  (-1)^{{\bf K}_1^th^{-1} Q(h^t)^{-1}{\bf K}_2} \ = \ (-1)^{{\bf K}_1^t Q{\bf K}_2} (-1)^{{\bf K}_1^t(h^{-1}Q(h^t)^{-1} -Q){\bf K}_2}\;. 
 \ee   
The failure of the cocycle to be $O(d,d,\mathbb{Z})$ invariant is encoded in the exponent of the second factor, 
which we compute with the help of the group property $h^{-1}=\eta^{-1}h^t \eta$ to be 
 \be
 \begin{split}
  {\cal A}(h) \ := \ h^{-1}Q(h^t)^{-1} -Q
  \ = \ \begin{pmatrix}   b^td & b^tc \\
  a^td-1 & a^tc  \end{pmatrix}
  \ = \ \begin{pmatrix}   b^td & b^tc \\
  -c^tb & a^tc  \end{pmatrix}
  \;. 
 \end{split}
 \ee
Here we used in the last step the $O(d,d)$ group properties, which also imply  that  $a^tc$ and $b^td$ 
are antisymmetric (see, e.g., sec.~4.1 in \cite{Kugo:1992md}). From this it follows that ${\cal A}(h)$ is antisymmetric, ${\cal A}^t=-{\cal A}$, 
so that it is determined by its upper-triangular part ${\cal A}_u$ via ${\cal A}={\cal A}_u-{\cal A}_u^t$. 
Let us record the transformation behavior of the cocycle factor: 
 \be\label{cocycleTRANS}
  C({\bf K}_1',{\bf K}_2') \ = \ C({\bf K}_1,{\bf K}_2)(-1)^{{\bf K}_1^t{\cal A}(h){\bf K}_2}\;. 
 \ee
 
Our goal is now to determine the sign factor in (\ref{ModTransformation}) by requiring that the 
$\bullet$ product transforms covariantly. For ease of notation we take the two factors $\phi^1$ and $\phi^2$ to be 
scalars so that $\rho(h)=1$. (This we can do without loss of generality since these finite-dimensional 
rotations are not relevant for the sign issues related to cocycles.) Therefore, the product (\ref{FourierModeProduct}) 
of two transformed fields reads 
 \be\label{FourierModeProductPrime}
 \begin{split}
  (\phi^{\prime 1}\bullet \phi^{\prime 2})_{{\bf K}'}
  &= \sum_{{\bf P}'\in \mathbb{Z}^{2d}} \delta_{{\bf P}'\cdot{\bf K}',0}\, C({\bf P}', {\bf K}') \,
  \phi_{{\bf P}'}^{\prime 1}\cdot \phi_{{\bf K}'-{\bf P}'}^{\prime 2}\\
  &=\sum_{{\bf P}\in \mathbb{Z}^{2d}} \delta_{{\bf P}\cdot{\bf K},0}\, C({\bf P}, {\bf K}) \,
  (-1)^{{\bf P}^t{\cal A}(h) {\bf K}}(-1)^{F({\bf P}, h)} (-1)^{F({\bf K}-{\bf P},h)} 
  \phi_{{\bf P}}^1\cdot \phi_{{\bf K}-{\bf P}}^2  \;, 
 \end{split}
 \ee
where we used (\ref{ModTransformation}) and (\ref{cocycleTRANS}). Covariance requires according to (\ref{ModTransformation})  
that this is equal to 
 \be
   (\phi^{\prime 1}\bullet \phi^{\prime 2})_{{\bf K}'} = (-1)^{F({\bf K}, h)} (\phi^1\bullet \phi^2)_{\bf K}\;, 
 \ee
 which is satisfied if 
  \be
   {\bf P}^t\,{\cal A}(h)\,{\bf K} = F({\bf K}, h)+ F({\bf P}, h) + F({\bf K}-{\bf P},h) \;\; {\rm mod} \; 2\;. 
  \ee
 We now prove that this relation holds for 
  \be\label{finalF}
   F({\bf K},h)  \equiv  {\bf K}^t\,{\cal A}_{u}(h) \,{\bf K}\;,
  \ee
 where ${\cal A}_u$ denotes  the upper-triangular part of ${\cal A}$. 
 This follows by a direct computation: 
 \be
 \begin{split}
 F({\bf K}, h)+ F({\bf P}, h) + F({\bf K}-{\bf P},h) &= {\bf K}^t{\cal A}_u(h){\bf K} 
 +{\bf P}^t{\cal A}_u(h){\bf P} +  ({\bf K}-{\bf P})^t{\cal A}_u(h)({\bf K}-{\bf P}) \\
 &=  {\bf P}^t\left({\cal A}_u(h) -  {\cal A}_{u}^t(h)\right) {\bf K}  \;\; {\rm mod} \; 2 \\
 &= {\bf P}^t \,{\cal A}_{u}(h)\,{\bf K}\;, 
 \end{split} 
 \ee   
 where we used repeatedly that this only needs to hold mod 2. 
 This completes the proof that under (\ref{ModTransformation}) with (\ref{finalF}) the 
 modified $\bullet$ product transforms covariantly. More generally it follows that an $O(d,d)$ singlet that is 
 built with any number of $\bullet$ products still transforms `covariantly' in this new sense. 
 As a consequence, any action integral based on such an $O(d,d)$ singlet is still $O(d,d,\mathbb{Z})$ invariant, 
 because the integral picks out the zero mode, c.f.~(\ref{zeromodeInt}), for which (\ref{finalF}) vanishes. 
 The modified $O(d,d,\mathbb{Z})$ transformation then reduces to the standard transformation (\ref{NaiveTransformation}) 
 under which the zero mode of a scalar is invariant.

\section{Decompositions}
In this section we decompose the fundamental fields of double field theory, 
specialized to the FRW backgrounds of sec.~\ref{FRWbackgrounds}, 
into irreducible components by performing a scalar-vector-tensor (SVT) decomposition. 
This allows us to 
express the quadratic double field theory action in terms of gauge invariant variables.  
This analysis extends what was already done for fluctuations with respect to a flat background in \cite{Chiaffrino:2020akd}.

\subsection{Scalar-vector-tensor decomposition}

We now take the backgrounds to be of the form (\ref{FRWdoubledFrame}) with scale factor $a(t)$. 
According to (\ref{Ltensorcomp}) we then have 
$L_{a}{}^b=L_{\bar{a}}{}^{\bar{b}}=0$,  so that the covariant derivatives defined in (\ref{def_bkgrd_nablabar}) simplify: 
\begin{align}
\begin{split}
\nabla_t \mathcal{V}_a = 
\dot{\mathcal{V}}_a \,, \qquad 
\nabla_t \mathcal{V}_{\bar{a}}  =  
\dot{\mathcal{V}}_{\bar{a}}\,. 
\end{split}
\end{align} 
As for double field theory on flat space we also have the spatial Laplacians that are related by the level-matching 
constraint as follows: 
\begin{equation}\label{DeltaReminder}
\Delta \equiv 
2 g^{ab}D_a D_b 
= 2g^{\bar{a}\bar{b}} D_{\bar{a}}D_{\bar{b}}  \,. 
\end{equation}
In contrast to flat space there is a  new differential operator based on the invariant tensor (\ref{mixedtensorsg}):
\begin{equation}
\label{DefDiamond}
\Diamond \equiv 
2 g_{a}{}^{\bar{b}} D^{a}D_{\bar{b}} = 2 g_{\bar{b}}{}^a D_a D^{\bar{b}}  \,, 
\end{equation} 
which satisfies  the commutation relations:
\begin{equation}
\begin{split}
[\nabla_t , \Delta ] & = 2 H \Diamond\,,\\
[\nabla_t, \Diamond ] & =   2 H \Delta  \,.\\
\end{split}
\end{equation}
Explicitly, these two Laplace-type operators are   given by 
 \be
 \begin{split}
  \Delta &= a^{-2}(t) \partial^2  + a^2(t) \tilde{\partial}^2\;,  \\
  \Diamond &=   - a^{-2}(t)\partial^2  +   a^2(t) \tilde{\partial}^2 \;,  
 \end{split}
 \ee
where 
 \be\label{delsquares} 
   \partial^2\equiv\partial^i\partial_i\equiv\delta^{ij}\partial_i\partial_j\;, 
   \qquad \tilde{\partial}^2\equiv \tilde{\partial}^i\tilde{\partial}_i
   \equiv \delta_{ij}\tilde{\partial}^i\tilde{\partial}^{j}\;, 
 \ee  
 are the independent spatial (Euclidean) Laplacians of the doubled space.  

At this stage a comment is in order regarding our conventions for raising and lowering indices. 
In the general frame formulation of double field theory the tangent space metric ${\cal G}_{AB}$ 
of signature $(d,d)$ is used to raise and lower flat indices. For the FRW backgrounds to be used here 
this metric takes the form (\ref{FRWtangentspace}) in terms of the positive-definite (Euclidean) metrics 
$g_{ab}$ and ${g}_{\bar{a}\bar{b}}$, respectively, which will be used from now on to raise and lower indices 
(as alluded to after (\ref{flatMetrics})). Since ${\cal G}_{ab}=-g_{ab}$ this amounts to a change in convention.
To be definite, let us take the elementary fields of the theory to be given by 
\be\label{generalFields}
  \phi\,, \quad \varphi\;, \quad h_{a\bar{b}}\;, \quad {\cal A}_{a}\;, \quad {\cal A}_{\bar{a}}\;, 
 \ee
 i.e., these fields are considered to be metric-independent, as are the differential 
 operators $\partial_a$, $\tilde{\partial}_a$, etc., with lower indices. Any expression involving 
 these objects with an upper index is then interpreted to 
 mean that the index is raised with $g^{ab}$ or  $g^{\bar{a}\bar{b}}$, respectively. 
 This change of convention leads to sign changes but has desirable internal consistency 
 properties. For instance, the Laplacians in (\ref{delsquares}) are then also writable as $\partial^2=\partial^a\partial_a=\partial^{\bar{a}}\partial_{\bar{a}}$ 
 and $\tilde{\partial}^2=\tilde{\partial}^a\tilde{\partial}_a=\tilde{\partial}^{\bar{a}}\tilde{\partial}_{\bar{a}}$, 
 as follows quickly by 
 recalling $\partial_a=e_{a}{}^{i}\partial_{i}$, $\tilde{\partial}_a=e_{a}{}^{i}\tilde{\partial}_i$, 
and similarly for barred indices. 
As a consequence, in all formulas to follow we may freely raise and lower indices without having 
to worry about sign factors. 


After this digression, we turn to the problem of 
decomposing   the complete list of fields (\ref{generalFields}) 
 into `irreducible' components by performing a scalar-vector-tensor (SVT) decomposition. 
For the ${\cal A}$ fields we write 
 \be\label{ADecomposition}
  \begin{split}
   {\cal A}_a &= A_a + \partial_a A + \tilde{\partial}_a  {\tilde{A}}  \;, \\
   {\cal A}_{\bar{a}} &=A_{\bar{a}} + \partial_{\bar{a}} \bar{A} +\tilde{\partial}_{\bar{a}}\bar{\tilde{A}} \;. 
  \end{split}
 \ee 
Here we see the first instance of an important novelty of a genuinely doubled 
field theory on cosmological backgrounds: the SVT decomposition of a vector yields, compared to 
standard gravity, an additional scalar mode, corresponding to the possibility of subtracting the divergence 
with respect to winding derivatives, in addition to ordinary derivatives. Correspondingly, the remaining vector mode
is now divergence-free (transverse) with respect to both derivatives: 
 \be\label{AConstraints}
  \partial^a A_a = \tilde{\partial}^a A_{a} = \partial^{\bar{a}} A_{\bar{a}}= \tilde{\partial}^{\bar{a}} A_{\bar{a}}=0\;. 
 \ee 
Thus, each of the transverse vectors has $d-2$ degrees of freedom. 
The logic here is that one imposes as many constraints as possible  on the remaining vector or tensor mode. 
This ultimately guarantees the complete decoupling among  tensor, vector and scalar modes. 
Turning then to the tensor field $h_{a\bar{b}}$ we 
postulate the decomposition 
 \be\label{hDecomposition}
  h_{a\bar{b}} =  \widehat{h}_{a\bar{b}} + g_{a\bar{b}} E    + \partial_a B_{\bar{b}} - \partial_{\bar{b}} B_a + \tilde{\partial}_{a} \tilde{B}_{\bar{b}} - \tilde{\partial}_{\bar{b}} \tilde{B}_a + \partial_a \partial_{\bar{b}} C + \tilde{\partial}_a \tilde{\partial}_{\bar{b}} \tilde{C} + \partial_{a}\tilde{\partial}_{\bar{b}} D + \tilde{\partial}_{a}\partial_{\bar{b}} \tilde{D}  \;, 
 \ee
with now five  independent scalar modes and four independent vector  modes, 
which are subject to the constraints analogous to (\ref{AConstraints}), 
i.e., every divergence vanishes: 
\begin{equation}
\partial^a B_a = \tilde{\partial}^a B_a = \partial^{\bar{a}} B_{\bar{a}} =\tilde{\partial}^{\bar{a}} B_{\bar{a}} = 0 \,, \qquad \partial^a \tilde{B}_a = \tilde{\partial}^a \tilde{B}_a = \partial^{\bar{a}} \tilde{B}_{\bar{a}} =\tilde{\partial}^{\bar{a}} \tilde{B}_{\bar{a}} = 0 \,.
\end{equation}
Similarly, the irreducible  tensor mode obeys 
\begin{equation}\label{tensorconstraints} 
\begin{split}
Q_{a} & \equiv  \partial^{\bar{b}} \widehat{h}_{a\bar{b}}=0 \,, \qquad Q_{\bar{b}} \equiv \partial^a  \widehat{h}_{a\bar{b}}=0\,,\\
\tilde{Q}_{a} & \equiv \tilde{\partial}^{\bar{b}} \widehat{h}_{a\bar{b}}=0 \,, \qquad \tilde{Q}_{\bar{b}}  \equiv \tilde{\partial}^a \widehat{h}_{a\bar{b}} =0\,, \qquad Q \equiv g_{\bar{b}}{}^a \widehat{h}_{a}{}^{\bar{b}}  =0 \,. 
\end{split}
\end{equation}
These constraints are not independent but rather subject to 
\begin{equation}
\begin{split}
\label{constraints-on-h-hat}
\partial^a Q_a & =\partial^{\bar{b}} Q_{\bar{b}} \,,\qquad 
\tilde{\partial}^a Q_a  = \partial^{\bar{b}} \tilde{Q}_{\bar{b}} \,,\\
\partial^a \tilde{Q}_a & = \tilde{\partial}^{\bar{b}} Q_{\bar{b}} \,,\qquad 
\tilde{\partial}^{a} \tilde{Q}_{a}  = \tilde{\partial}^{\bar{b}}\tilde{Q}_{\bar{b}} \,.
\end{split}
\end{equation}     
Let us verify that  ${h}_{a\bar{b}}$ so written encodes the right number of (off-shell) degrees of freedom. As for $A$, the vector  modes  $B_a$, $B_{\bar{a}}$, $\tilde{B}_a$, $\tilde{B}_{\bar{a}}$ together encode 
$4(d-2)$ degrees of freedom. The number of components of $\widehat{h}_{a\bar{b}}$ is $d^2$ minus the number of 
constraints. Since the constraints (\ref{tensorconstraints})  in turn are subject to  (\ref{constraints-on-h-hat}) we have 
$4d+1-4=4d-3$ independent constraints, so that $\widehat{h}_{a\bar{b}}$ carries $d^2 -4d+3 $ degrees of freedom. 
In total, together with the five scalar modes $E,C,\tilde{C},D,\tilde{D}$ and  the $4(d-2)$ vector modes, 
the irreducible components  carry $4(d-2) + d^2 -4d + 3 + 5 = d^2$ degrees of freedom, as it should be.

At this stage it is appropriate to briefly discuss the counting of degrees of freedom done above and to relate 
it to the familiar counting in, say, four spacetime dimensions (for which $d=3$). Consider, for instance, 
the Fourier expansion of a vector mode in (\ref{ADecomposition}): 
 \be
  A_a(x,\tilde{x}) = \sum_{k,\tilde{k}}A_{a}(k,\tilde{k}) e^{i(k\cdot x+\tilde{k}\cdot \tilde{x})}\;. 
 \ee
As before we assume that the Fourier modes $A_{a}(k,\tilde{k})$ are only non-zero provided $k\cdot\tilde{k}=0$, 
so that the weak constraint (level-matching constraint) is obeyed. The two constraints in (\ref{AConstraints}) yield
 \be
  k^a A_{a}(k,\tilde{k}) = \tilde{k}^aA_{a}(k,\tilde{k})=0 \;, 
 \ee 
which implies that among the three components of $A_{a}$ (for $d=3$) generically only one survives. For instance, 
for $k=(0,0,1)$ and $\tilde{k}=(0,1,0)$, for which $k\cdot\tilde{k}=0$, the constraints imply $A_2=A_3=0$, 
so that only $A_1$ remains as a physical degree of freedom. So what happened to the familiar two polarizations of 
a spin-1 vector mode in four dimensions? These spin-1 modes are still present for the special case that the individual modes
carry only momentum or only winding, say in the form 
 \be
  A_a(x,\tilde{x}) = \sum_{k}A_{a}(k) e^{i k\cdot x}
  +\sum_{\tilde{k}}\tilde{A}_{a}(\tilde{k}) e^{i \tilde{k}\cdot \tilde{x}}\;, 
 \ee
for which the level-matching constraint is trivially satisfied. Then for each of the two terms one of the constraints 
(\ref{AConstraints})  trivializes, so that the corresponding vector mode carries the expected two polarizations. 
Therefore, if we consider vector modes that live only in $x$-space, or vector modes that live only in $\tilde{x}$-space, 
they do carry the familiar two polarizations (effectively eliminating the additional scalar modes). It is only for modes 
that both carry genuine momentum and winding that the degrees of freedom organize differently. 
Similar remarks apply to the tensor modes. Indeed, naively 
$\widehat{h}_{a\bar{b}}$ carries zero degrees of freedom in four dimensions, but if we consider tensor modes 
that live only in $x$-space, or only in $\tilde{x}$-space, some of the constraints trivialize, so that they do 
carry the two polarizations of a spin-2 mode.

Returning to our discussion of the SVT decomposition, let us verify that these  decompositions exist by proving that the SVT components satisfying 
the appropriate constraints can always 
be defined from the given fields (\ref{generalFields}).  More precisely, this is the case if 
$\partial^2$ and $\tilde{\partial}^2$ are invertible operators, as we will assume in the following. 
For instance,  
the scalar modes of $h_{a\bar{b}}$ 
can be expressed in terms of the original fields as: 
  \begin{equation}
\begin{split}
\label{scalarintermsofh}
C & = \frac{d-1}{d-2} \partial^{-4} ( \partial^a \partial^{\bar{b}} h_{a\bar{b}} ) - \frac{1}{d-2}  \partial^{-2}  \big( g_{\bar{b}}{}^a h_{a}{}^{\bar{b}}   - \tilde{\partial}^{-2} ( \tilde{\partial}^a \tilde{\partial}^{\bar{b}} h_{a\bar{b}} ) \big)  \,,\\ 
\tilde{C}& = \frac{d-1}{d-2}  \tilde{\partial}^{-4} ( \tilde{\partial}^{a}\tilde{\partial}^{\bar{b}} h_{a\bar{b}} ) -  \frac{1}{d-2} \tilde{\partial}^{-2} \big( g_{\bar{b}}{}^a h_{a}{}^{\bar{b}} - \partial^{-2} ( \partial^a \partial^{\bar{b}} h_{a\bar{b}}) \big)  \,,\\
D & = \partial^{-2} \tilde{\partial}^{-2} ( \partial^a \tilde{\partial}^{\bar{b}} h_{a\bar{b}} ) \,,\qquad 
\tilde{D} = \tilde{\partial}^{-2} \partial^{-2} ( \tilde{\partial}^a \partial^{\bar{b}} h_{a\bar{b}} ) \,, \\
E  &  =  \frac{1}{d-2} \big( g_{\bar{b}}{}^a h_{a}{}^{\bar{b}}  - \partial^{-2} \partial^a \partial^{\bar{b}} h_{a\bar{b}} - \tilde{\partial}^{-2} \tilde{\partial}^a \tilde{\partial}^{\bar{b}} h_{a\bar{b} } \big)   \,. 
\end{split}
\end{equation} 
The vector modes of $h_{a\bar{b}}$ in turn can be defined as 
\begin{equation}
\begin{split} 
\label{vectormodesintermsofh}
B_a & = - \partial^{-2} \big( \partial^{\bar{b}} h_{a\bar{b}} -  g_{a\bar{b}} \partial^{\bar{b}} E \big)  + \partial_a C + \tilde{\partial}_a \tilde{D}  \,, \\ 
B_{\bar{b}}   &= \partial^{-2} \big( \partial^a h_{a\bar{b}} - g_{a\bar{b}} \partial^a E \big) - \partial_{\bar{b}} C - \tilde{\partial}_{\bar{b}} D \,,\\
\tilde{B}_a & =   - \tilde{\partial}^{-2} \big( \tilde{\partial}^{\bar{b}} h_{a\bar{b}} - g_{a\bar{b}} \tilde{\partial}^{\bar{b}} E \big) + \tilde{\partial}_a \tilde{C} + \partial_a \tilde{D} \,,\\
\tilde{B}_{\bar{b}}  &= \tilde{\partial}^{-2} \big( \tilde{\partial}^{a} h_{a\bar{b}} - g_{a\bar{b}} \tilde{\partial}^{a} E \big) - \tilde{\partial}_{\bar{b}} \tilde{C} - \partial_{\bar{b}} D  \,,
\end{split}
\end{equation}
where  one should view the scalar modes in here as defined in terms of $h_{a\bar{b}}$ via (\ref{scalarintermsofh}). 
Finally, $\widehat{h}_{a\bar{b}}$ can then be defined by inserting these scalar and vector modes (\ref{scalarintermsofh}) and (\ref{vectormodesintermsofh}) into (\ref{hDecomposition}) 
and solving for  $\widehat{h}_{a\bar{b}}$. 
Similarly, the scalar and vector modes of the vector fields $\mathcal{A}_a$ and  $\mathcal{A}_{\bar{a}}$ 
can be defined in terms of these fields as 
\begin{equation}
\begin{split}
A & = \partial^{-2} ( \partial^a \mathcal{A}_a ) \,,\qquad
\tilde{A}  = \tilde{\partial}^2 ( \tilde{\partial}^a \mathcal{A}_a ) \,,\qquad 
\bar{A}  = \partial^{-2} ( \partial^{\bar{a}} \mathcal{A}_{\bar{a}} ) \,,\qquad 
\bar{\tilde{A}}  = \tilde{\partial}^{-2} ( \tilde{\partial}^{\bar{a}} \mathcal{A}_{\bar{a}} )  \,,\\
A_a & = \mathcal{A}_a - \partial_a \big(\partial^{-2} ( \partial^b \mathcal{A}_b)\big) - \tilde{\partial}_a \big(\partial^{-2} ( \partial^{\bar{b}} \mathcal{A}_{\bar{b}} )\big)  \,,\\
A_{\bar{a}} & = \mathcal{A}_{\bar{a}} - \partial_{\bar{a}} \big( \tilde{\partial}^{-2} ( \tilde{\partial}^{\bar{b}} \mathcal{A}_{\bar{b}} )\big)  - \tilde{\partial}_{\bar{a}} \big(\tilde{\partial}^{-2} ( \tilde{\partial}^{\bar{b}} \mathcal{A}_{\bar{b}} ) \big) \,.
\end{split}
\end{equation}

We will next determine the gauge transformations of the SVT components.  To this end we decompose the gauge parameters $\xi_a$ and $\xi_{\bar{a}}$ into scalar and divergenceless vector components: 
\begin{equation}
\xi_a =  \zeta_a + \partial_a \lambda + \tilde{\partial}_a \chi  \,, \qquad 
\xi_{\bar{a}} = \zeta_{\bar{a}} + \partial_{\bar{a}} \bar{\lambda} + \tilde{\partial}_{\bar{a}}\bar{\chi} \,, 
\end{equation}
where $ \partial^a \zeta_a= \tilde{\partial}^a \zeta_a= \partial^{\bar{a}}\zeta_{\bar{a}}=\tilde{\partial}^{\bar{a}}\zeta_{\bar{a}}=0$. 
With this decomposition, the gauge transformations (\ref{quadrgauge1234}) become:
\begin{equation}
\begin{split}
\label{GT delta habbar}
\delta h_{a\bar{b}} 
 & = \frac{1}{\sqrt{2}}  \big( a \tilde{\partial}_a \zeta_{\bar{b}} - a^{-1} \partial_a \zeta_{\bar{b}}   - a \tilde{\partial}_{\bar{b}} \zeta_a - a^{-1} \partial_{\bar{b}} \zeta_a \big) \\&
 +  \frac{1}{\sqrt{2}} \big( \tilde{\partial}_a \partial_{\bar{b}} (a\bar{\lambda}-a^{-1} \chi )
  +  a \tilde{\partial}_a\tilde{\partial}_{\bar{b}} ( \bar{\chi}- \chi) - a^{-1} \partial_a \partial_{\bar{b}} (\lambda+\bar{\lambda}) -  \partial_a \tilde{\partial}_{\bar{b}} ( a^{-1}\bar{\chi}+ a \lambda) 
  \big)  - \xi^0 L_{a\bar{b}}  \,,\\
  \delta \phi &= \dot{\xi}^0 \,,\\
  \delta \varphi & =   \dot{\Phi} \xi^0
+ \frac{1}{2\sqrt{2}} \big( -a^{-1} \partial^2 ( \lambda + \bar{\lambda}) + a \tilde{\partial}^2 ( \chi - \bar{\chi}) \big)   \,,\\
\delta \mathcal{A}_a & = \dot{\zeta}_a - L_{a}{}^{\bar{b}} \zeta_{\bar{b}} + \partial_a \bigg(\dot{\lambda} - H \bar{\lambda}- \frac{a^{-1}}{\sqrt{2}}( \tilde{\xi}_0 + \xi^0) \bigg) + \tilde{\partial}_a\bigg ( \dot{\chi} - H \bar{\chi} + \frac{a}{\sqrt{2}} (  \tilde{\xi}_0 + \xi^0)  \bigg)  \,,\\
\delta \mathcal{A}_{\bar{a}} & = \dot{\zeta}_{\bar{a}} - L_{\bar{a}}{}^b \zeta_b + \partial_{\bar{a}} \bigg(\dot{\bar{\lambda}} - H \lambda  + \frac{a^{-1}}{\sqrt{2}} ( \tilde{\xi}_0 - \xi^0) \bigg)  + \tilde{\partial}_{\bar{a}} \bigg( \dot{\bar{\chi}} - H \chi+ \frac{a}{\sqrt{2}} ( \tilde{\xi}_0 - \xi^0)  \bigg) \,. 
 \end{split}
\end{equation}
Comparing this with (\ref{ADecomposition}), (\ref{hDecomposition}) we 
can then read off the gauge transformation of the SVT components of the fields: 
\begin{equation}
\begin{split}
\label{svtcomponnentsofh}
\delta\widehat{h}_{a\bar{b}} &= 0 \,,  \qquad \delta E  = - H  \xi^0   \,,    \\
\delta B_a  & = \frac{a^{-1}}{\sqrt{2}} \zeta_a \,,\qquad 
\delta B_{\bar{a}}  = -\frac{a^{-1}}{\sqrt{2}} \zeta_{\bar{a}} \,,\qquad 
\delta \tilde{B}_a  = \frac{a}{\sqrt{2}} \zeta_{a} \,,\qquad 
\delta \tilde{B}_{\bar{a}}  = \frac{a}{\sqrt{2}} \zeta_{\bar{a}}\,,  \\
\delta C & = -\frac{ a^{-1} }{\sqrt{2}} ( \lambda + \bar{\lambda} ) \,, \qquad \qquad 
\delta \tilde{C} = \frac{a}{\sqrt{2}}  ( \bar{\chi } - \chi ) \,, \quad  \\
\delta D  & = \frac{1}{\sqrt{2}} ( - a^{-1} \bar{\chi} - a \lambda ) \,,\qquad 
\delta \tilde{D} = \frac{1}{\sqrt{2}} ( a \bar{\lambda} - a^{-1} \chi ) \,,  \\
\delta A_a & = \dot{\zeta}_a -  L_{a} {}^{\bar{b}}\zeta_{\bar{b}} \,,\qquad \qquad 
\delta A_{\bar{a}}  = \dot{\zeta}_{\bar{a}} -  L_{\bar{a}}{}^b \zeta_{b} \,, \\ 
\delta A & = \dot{\lambda} - H \bar{\lambda}   - \frac{a^{-1}}{\sqrt{2}} ( \tilde{\xi}_0 + \xi^0)   \,,\qquad 
\delta \tilde{A} = \dot{\chi}  - H \bar{\chi}   + \frac{a}{\sqrt{2}} ( \tilde{\xi}_0 + \xi^0) \,,\\
\delta \bar{A} &=  \dot{\bar{\lambda} }  -H \lambda + \frac{a^{-1}}{\sqrt{2}} ( \tilde{\xi}_0 - \xi^0) \,,\qquad 
\delta \bar{\tilde{A}} =\dot{\bar{\chi}} - H \chi   + \frac{a}{\sqrt{2}} ( \tilde{\xi}_0 - \xi^0) \,. 
\end{split}
\end{equation} 
Note that the tensor mode $\widehat{h}_{a\bar{b}}$ is gauge invariant.

\subsection{Gauge fixing} 

In order to verify that the number of gauge independent fields is as expected (i.e.~equal to the number of off-shell 
degrees of freedom minus the number of gauge redundancies) we can impose simple gauge fixing conditions, 
as we do in the following. 

From the second line in (\ref{svtcomponnentsofh}) we see that $\zeta_a$ and $\zeta_{\bar{a}}$ can be used 
to gauge fix two vectors to zero, e.g., 
 \be
  \tilde{B}_{a} = \tilde{B}_{\bar{a}}=0\;. 
 \ee
This fixes the gauge invariance under $\zeta_a$, $\zeta_{\bar{a}}$ completely and does not require any  
compensating gauge transformations. Next, we observe  with the last two lines in (\ref{svtcomponnentsofh}) 
that $\xi^0$ and $\tilde{\xi}_0$ can be used to gauge away two scalars, say:
\be
  \tilde{A} = \bar{\tilde{A}}=0\;. 
\ee
This again fixes the gauge invariance under $\xi^0$, $\tilde{\xi}_0$ completely, but now we require compensating 
gauge transformations that determine these parameters in terms of the remaining scalar gauge parameters: 
 \be\label{compesatingxhis}
  \frac{a}{\sqrt{2}} ( \tilde{\xi}_0 + \xi^0) = -\dot{\chi}  + H \bar{\chi} \;, \qquad
  \frac{a}{\sqrt{2}} ( \tilde{\xi}_0 - \xi^0)  = -\dot{\bar{\chi}} + H \chi  \;. 
 \ee
 We are now left with four scalar gauge parameters ($\lambda$, $\chi$, $\bar{\lambda}$, $\bar{\chi}$), 
 so naively we would expect that we can gauge away four more scalar field modes, but it turns out 
 that, due to the gauge for gauge symmetry, only three more scalar modes can be gauged away. 
 Let us pick the gauge that, say,  
  \be
   \tilde{C}= D=\tilde{D}=0\;. 
  \ee 
The first condition fixes, say, the parameter $\bar{\chi}=\chi$. The second gauge condition can then 
be achieved by means of $\lambda$, which in turn fixes the compensating gauge transformation 
to be $\lambda=-a^{-2}\bar{\chi}=-a^{-2}\chi$. 
Finally, the third gauge condition can 
be achieved by means of $\bar{\lambda}$, which in turn fixes the compensating gauge transformation 
to be $\bar{\lambda}=a^{-2}\chi$. In total we have reduced the gauge redundancy 
to one, with independent parameter $\chi$ and compensating transformations in terms of $\chi$: 
 \be\label{compensatingchis}
  \bar{\chi}=\chi\;, \qquad \lambda =-a^{-2}\chi\;, \qquad \bar{\lambda}=a^{-2}\chi\;. 
 \ee
However, from the third line of (\ref{svtcomponnentsofh}) we then infer that $C$ is gauge invariant and thus cannot 
be set to zero. Similarly, using (\ref{compensatingchis}) in (\ref{compesatingxhis}) yields 
 \be
  \tilde{\xi}_0 = -\sqrt{2} a^{-1} (\dot{\chi}-H{\chi})\;, \qquad \xi^0=0\;, 
 \ee
and with this it follows  that the other remaining scalar modes ($E$, $A$ and $\bar{A}$) 
are all gauge invariant and can thus not be gauged away. Thus,  the apparent remaining 
parameter $\chi$ in fact does not act at all on the remaining fields. This is just a consequence of  there 
being a scalar gauge-for-gauge symmetry. Thus, we have fixed the gauge redundancy completely. 

Summarizing, the gauge independent fields can be chosen to be, for instance: 
  \begin{itemize}
   \item tensor modes: $\qquad$ $\widehat{h}_{a\bar{b}}$ $\qquad \qquad\qquad\qquad\qquad \;\,$ 
   $[d^2-4d +3]$\,,
   \item vector modes: $\qquad$ $A_{a}$\,, $A_{\bar{a}}$\,, $B_{a}$\,, $B_{\bar{a}}$
   $\qquad \qquad\;\;\,$ $[4(d-2)]$\,,
   \item scalar modes:  $\qquad$ $E$\,, $C$\,, $A$\,, $\bar{A}$\,, $\phi$, $\varphi$
    $\qquad \qquad$ $[6] $\;,  
  \end{itemize} 
where we also included the two scalar modes $\phi$, $\varphi$ that were present from the beginning. 
We displayed  in parenthesis the number of degrees of freedom, which adds to $d^2+1$. 
This is equal to the total number of off-shell degrees of freedom of the fields (\ref{generalFields}), given by 
$d^2+2d+2$, minus the number of gauge redundancies for parameters $\xi_{a}$, $\xi_{\bar{a}}$, $\tilde{\xi}_0$, $\xi^0$, 
which is given by $2d+1$ (taking into account the one gauge-for-gauge redundancy).

\subsection{Gauge invariant variables}

We now aim to rewrite the field variables and the theory in terms of gauge invariant combinations of the SVT components. 
This is essentially equivalent to fixing a gauge in the sense that the resulting gauge invariant variables are subject to constraints that are formally 
identical to gauge fixing conditions. A gauge invariant formulation may always be reconstructed from a gauge 
fixed one. 
We have already noted that $\widehat{h}_{a\bar{b}}$ gauge invariant. 
In addition, one can build  the following gauge invariant variables:
\begin{equation}\label{NewBardeens}
\begin{split}
\widehat{\phi} &  = \phi + \frac{1}{2\sqrt{2}}\nabla_t   \big(  a ( A + \bar{A} + \sqrt{2}  a \dot{C}   )  
- a^{-1} ( \tilde{A} - \bar{\tilde{A}} + \sqrt{2}  a^{-1} \dot{\tilde{C}}    )  \big)   \,,\\ 
\widehat{A}_a & = A_a - \frac{1}{\sqrt{2}} \nabla_t ( a B_a + a^{-1} \tilde{B}_a) - \frac{1}{\sqrt{2}} L_{a}{}^{\bar{b}} ( a B_{\bar{b}} - a^{-1} \tilde{B}_{\bar{b}} ) \,, \\ 
\widehat{A}_{\bar{a}} & = A_{\bar{a}}  + \frac{1}{\sqrt{2}} \nabla_t ( a B_{\bar{a}} - a^{-1} \tilde{B}_{\bar{a}} ) + \frac{1}{\sqrt{2}} L_{\bar{a}}{}^{b} ( a B_b + a^{-1} \tilde{B}_b ) \,,\\ 
\widehat A & =   a( A + \bar{A} + \sqrt{2}a\dot{C} ) + a^{-1} ( \tilde{A} - \bar{\tilde{A}} + \sqrt{2} a^{-1} \dot{\tilde{C}} )    \,, \\ 
\widehat{\tilde{A}} & = 
 a ( A - \bar{A} - \sqrt{2} a\dot{C} -2 \sqrt{2}a H C ) + a^{-1} ( \bar{\tilde{A}} + \tilde{A} + \sqrt{2} a^{-1} \dot{\tilde{C}} - 2\sqrt{2} a^{-1} H\tilde{C}) + 2 \sqrt{2}\dot{D}   \,,\\  
\widehat{\bar{A}} & =  a ( \bar{A} - A - \sqrt{2} a \dot{C} - 2 \sqrt{2} a H C ) - a^{-1} ( \bar{\tilde{A}} + \tilde{A} - \sqrt{2} a^{-1} \dot{\tilde{C}}+2 \sqrt{2} a^{-1} H \tilde{C} ) - 2 \sqrt{2} \dot{\tilde{D}}\,,\\
\widehat{C}& = D- \tilde{D} - a^2 C + a^{-2} \tilde{C} \,,\\ 
\widehat{B}_a & = \frac{1}{2}( a B_a - a^{-1} \tilde{B}_a ) \,,\\
\widehat{B}_{\bar{a}} & =\frac{1}{2}(  a B_{\bar{a}} +  a^{-1} \tilde{B}_{\bar{a}} ) \,,\\
\widehat{\varphi} & = \varphi + \frac{1}{2\sqrt{2}}    \dot{\Phi} \big(  a ( A + \bar{A} + \sqrt{2}  a \dot{C}   )  
- a^{-1} ( \tilde{A} - \bar{\tilde{A}} + \sqrt{2}  a^{-1} \dot{\tilde{C}}    ) \big)   + \frac{1}{2} ( -\partial^2 C + \tilde{\partial}^2 \tilde{C} )  \,,\\
 \widehat{E} & =  E - \frac{1}{2\sqrt{2}}  H  \big ( a ( A + \bar{A} + \sqrt{2}  a \dot{C}   )  
- a^{-1} ( \tilde{A} - \bar{\tilde{A}} + \sqrt{2}  a^{-1} \dot{\tilde{C}}    )\big )   \,, 
\end{split}
\end{equation} 
as follows with simple computations using (\ref{svtcomponnentsofh}).
These gauge invariant variables are of course not unique, as any linear combination of gauge invariant fields 
is also gauge invariant. Moreover, there is one  relation among them: 
the scalars $\widehat{\tilde{A}}$ and $\widehat{\bar{A}}$ satisfy 
\begin{equation}
\widehat{\tilde{A}} + \widehat{\bar{A}}   =  2\sqrt{2}\dot{\widehat{C}}\,. 
\end{equation}

We next aim to rewrite the quadratic double field theory action directly in terms of  gauge invariant variables 
that are linear combinations of (\ref{NewBardeens}). 
Given that the latter take a rather complicated form, a priori this seems to be a difficult task that, however, is simplified 
 by the following trick: We use that the original fields can be expressed in terms of specific 
combinations of the  gauge 
invariant fields plus terms that take the form of an infinitesimal gauge transformation (\ref{quadrgauge1234}). 
Specifically, we claim that 
\begin{equation}\label{gaugeinvSPlit}
\begin{split}
h_{a\bar{b}} \ & =  \ \bar{h}_{a\bar{b}} + D_a F_{\bar{b}} - D_{\bar{b}}F_a   - F^0 L_{a\bar{b}} \,,\\
\mathcal{A}_a \ & = \ \bar{\mathcal{A}}_a  
+\dot{F}_a    - L_{a}{}^{\bar{b}} F_{\bar{b}}  
+ D_{a} \big( \tilde{F}_0 + F^{0} \big)\\ 
\mathcal{A}_{\bar{a}} \ & = \ \bar{\mathcal{A}}_{\bar{a}}  + \dot{F}_{\bar{a}} - L_{\bar{a}}{}^b F_b + D_{\bar{a}} ( \tilde{F}_0 - F^0) \,,\\ 
\varphi \ & = \ \bar{\varphi} + \dot{\Phi} F^0  - \tfrac{1}{2} D_a F^a - \tfrac{1}{2} D_{\bar{a}} F^{\bar{a}}\,,\\
\phi \ & = \ \bar{\phi}-\dot{F}^0 \,, 
\end{split}
\end{equation}
where the gauge invariant fields, denoted by a bar, are given by 
\begin{flalign}
\bar{h}_{a\bar{b}} 
&= \widehat{h}_{a\bar{b}} +(a^{-1}\partial_a+a\tilde{\partial}_a)\widehat{B}_{\bar{b}} 
-(a^{-1}\partial_{\bar{b}}-a\tilde{\partial}_{\bar{b}}) \widehat{B}_{a} +  \widehat{E} g_{a\bar{b}}\nonumber\\
&\quad +\frac{1}{2}(a^{-2} \partial_a\partial_{\bar{b}} +2\partial_a\tilde{\partial}_{\bar{b}} - 2\tilde{\partial}_{a}\partial_{\bar{b}}
-a^2 \tilde{\partial}_{a}\tilde{\partial}_{\bar{b}})\widehat{C} 
\,,\\
 \bar{\mathcal{A}}_a & =\widehat{\mathcal{A}}_a + \frac{a^{-1}}{4} \partial_a \big(   \widehat{A} - \widehat{\bar{A}}   +  2\sqrt{2} H \widehat{C}  \big) + \frac{a}{4} \tilde{\partial}_a \big(   \widehat{A} + \widehat{\tilde{A}}   + 2 \sqrt{2} H \widehat{C} \big)
 \,, \\ 
 \bar{\mathcal{A}}_{\bar{a}} & =
\widehat{\mathcal{A}}_{\bar{a}} +\frac{a^{-1}}{4} \partial_{\bar{a}} \big( \widehat{A} - \widehat{\tilde{A}}  + 2\sqrt{2}  H \widehat{C} \big) -\frac{a}{4} \tilde{\partial}_{\bar{a}} \big(    \widehat{A} + \widehat{\bar{A}}  + 2\sqrt{2}H \widehat{C}\big)\,,\\
\bar{\varphi}& =
\widehat{\varphi} +\frac{1}{4}\Delta \widehat{C} \,,\\
\bar{\phi}&  = \widehat{\phi}\,, 
\end{flalign} 
while the `effective gauge parameters', denoted by $F$, are given by 
\begin{equation}\label{Fgaugestuff}
\begin{split}
F^0 & = - \frac{1}{2\sqrt{2}}   \big( a ( A + \bar{A} + \sqrt{2}  a \dot{C}   )  
- a^{-1} ( \tilde{A} - \bar{\tilde{A}} + \sqrt{2}  a^{-1} \dot{\tilde{C}}    ) \big)  \,,  \\  
 F_0 &= \frac{1}{2\sqrt{2}} \big( a^{-1} ( \tilde{A} + \bar{\tilde{A}}   ) - a ( A - \bar{A} )\big) \,,\\
  F_a & =  \frac{1}{\sqrt{2}} ( a B_a + a^{-1} \tilde{B}_a )+ \frac{\sqrt{2}a^{-1}}{4}\partial_a (  -3 a^2 C - 2 \tilde{D} + a^{-2} \tilde{C} ) 
  + \frac{\sqrt{2} a}{4} \tilde{\partial}_a ( -3 a^{-2} \tilde{C} - 2 D +a^2 C )  \,,\\
  F_{\bar{a}}& =    - \frac{1}{\sqrt{2}}  (   a B_{\bar{a}}  - a^{-1} \tilde{B}_{\bar{a}} )
  + \frac{\sqrt{2}a^{-1}}{4} \partial_{\bar{a}}  ( -3   a^2 C  + 2 D  +  a^{-2} \tilde{C} ) 
  +\frac{\sqrt{2}a}{4}  \tilde{\partial}_{\bar{a}} ( 3a^{-2} \tilde{C} - 2 \tilde{D} -a^2 C    ) \,.   
 \end{split}
\end{equation}
These quantities transform under gauge transformations as 
 \be
 \begin{split} 
    \delta F^0 & = \xi^0 \;,\\
  \delta F_{0} &=\tilde{\xi}_0 - \dot{\eta}\;,\\
  \delta F_a &= \zeta_a + \partial_a\big(\lambda -\tfrac{1}{\sqrt{2}}a^{-1}\eta\big) +\tilde{\partial}_a
  \big(\chi+\tfrac{1}{\sqrt{2}}a\eta\big)  = \xi_a  -\tfrac{1}{\sqrt{2}} a^{-1}\partial_a\eta 
  + \tfrac{1}{\sqrt{2}}a \tilde{\partial}_a\eta\;, \\
  \delta F_{\bar{a}} &=\zeta_{\bar{a}} + \partial_{\bar{a}} \big( \bar{\lambda} + \tfrac{1}{\sqrt{2}} a^{-1}\eta  \big) + \tilde{\partial}_{\bar{a}}  \big(\bar{\chi}+ \tfrac{1}{\sqrt{2}} a \eta \big ) = \xi_{\bar{a}}  + \frac{1}{\sqrt{2}}  a^{-1} \partial_{\bar{a}}\eta +\frac{1}{\sqrt{2}} a \tilde{\partial}_{\bar{a}} \eta \;,\\
 \end{split}
 \ee
where the gauge-for-gauge parameter $\eta$ is given by 
 \be
  \eta=\frac{\sqrt{2}}{4} \left(a\lambda -a\bar{\lambda} -a^{-1}\chi -a^{-1}\bar{\chi}\right)\;. 
 \ee 
Using the expressions (\ref{Fgaugestuff})  and those for the gauge invariant variables (\ref{NewBardeens}) one may verify 
that (\ref{gaugeinvSPlit}) are just identities. These identities are very useful, however, since they decompose the fields 
into their gauge invariant parts plus terms of the `pure gauge form' (\ref{quadrgauge1234}).

We can now replace each appearance of a field in the quadratic action by its right-hand side in (\ref{gaugeinvSPlit}). 
Gauge invariance then implies that the pure gauge terms drop out, so that in effect we 
may simply replace each field by its gauge invariant version, i.e., we just put a bar on each field. 
Afterwards we use the expressions in terms of the SVT components, which in turn allows us to 
express each divergence, trace, etc., of a tensor or vector in terms of appropriate scalar modes, 
thereby achieving complete decoupling. 

In order to perform this computation 
it is convenient to first display the
quadratic  Lagrangian (\ref{SimpleActionQuadLagrangian}) in terms of the derivatives $\partial$ and $\tilde{\partial}$: 
\begin{equation} 
\begin{split}
\mathcal{L} = & \, n e^{-2\Phi} \bigg\{ 
- 4 \dot{\varphi}^2 + 8 \dot{\Phi} \phi \dot{\varphi}  +H(2 {\phi}+4  {\varphi}) g_{\bar{b}}{}^{a}\dot{h}_{a}{}^{\bar{b}} 
+ \dot{h}_{a\bar{b}} \dot{h}^{a\bar{b}} + 2 H^2 h^{a\bar{b}} h_{a\bar{b}} - 2 H^2 g_{a}{}^{\bar{c}} g_{\bar{d}}{}^b h^{a\bar{d}} h_{b\bar{c}} 
\\&
+ \frac{1}{\sqrt{2}} ( a \tilde{\partial}^{\bar{a}} \mathcal{A}_{\bar{a}} + a^{-1} \partial^{\bar{a}} \mathcal{A}_{\bar{a}}-a \tilde{\partial}^a \mathcal{A}_a + a^{-1} \partial^a \mathcal{A}_a  ) (- 4 \dot{\varphi} + 4 \dot{\Phi} \phi - 2 Hg_{\bar{b}}{}^{a} h_a{}^{\bar{b}} ) 
\\&
+ \frac{1}{\sqrt{2}} H  ( 4\varphi+ 2 \phi  ) ( - a\tilde{\partial}^{\bar{a}} \mathcal{A}_{\bar{a}} + a^{-1} \partial^{\bar{a}} \mathcal{A}_{\bar{a}} +a \tilde{\partial}^a \mathcal{A}_a+ a^{-1} \partial^a \mathcal{A}_a ) 
\\
&
- \frac{1}{2} \mathcal{A}_{\bar{a}} \Delta \mathcal{A}^{\bar{a}} - \frac{1}{2} \mathcal{A}^a \Delta \mathcal{A}_a - \frac{1}{2} ( a \tilde{\partial}^a \mathcal{A}_a - a^{-1} \partial^a \mathcal{A}_a )^2 - \frac{1}{2} ( a \tilde{\partial}^{\bar{a}} \mathcal{A}_{\bar{a}}+ a^{-1} \partial^{\bar{a}} \mathcal{A}_{\bar{a}} )^2 
\\&
+ \sqrt{2}\mathcal{A}^{\bar{b}} ( a \tilde{\partial}^a -a^{-1} \partial^a) \dot{h}_{a\bar{b}} 
-\sqrt{2}  \mathcal{A}^a ( a \tilde{\partial}^{\bar{b}} + a^{-1} \partial^{\bar{b}} ) \dot{h}_{a\bar{b}} 
-\sqrt{2} Hg_{b}{}^{\bar{d}} \mathcal{A}^b ( a \tilde{\partial}^a h_{a\bar{d}} - a^{-1} \partial^a h_{a\bar{d}} ) \\&
+   \sqrt{2} Hg_{\bar{b}}{}^c \mathcal{A}^{\bar{b}} ( a \tilde{\partial}^{\bar{a}} h_{c\bar{a}} + a^{-1} \partial^{\bar{a}} h_{c\bar{a}} )
 -  \sqrt{2} H \mathcal{A}^{\bar{b}} ( a \tilde{\partial}^a h_{a\bar{b}} + a^{-1} \partial^a h_{a\bar{b}} ) \\&
+                         \sqrt{2} H \mathcal{A}^a ( a \tilde{\partial}^{\bar{b}} h_{a\bar{b}} - a^{-1} \partial^{\bar{b}} h_{a\bar{b}} ) \\
& + 4 \phi \Delta \varphi - 4 \varphi \Delta \varphi +  ( 4\varphi- 2\phi) (a^{-2}\partial^a \partial^{\bar{b}} - \tilde{\partial}^a \partial^{\bar{b}} + \partial^a \tilde{\partial}^{\bar{b}} - a^2 \tilde{\partial}^a \tilde{\partial}^{\bar{b}} )h_{a\bar{b}} + h^{a\bar{b}} \Delta h_{a\bar{b}} \\&
+  (a^{-1}\partial^a -a \tilde{\partial}^a) h_{a}{}^{\bar{b}}  (a^{-1}\partial^c -a \tilde{\partial}^c )  h_{c\bar{b}} +   (a^{-1}\partial^{\bar{b}}+a \tilde{\partial}^{\bar{ b}} )h_{a\bar{b}}   (a^{-1}\partial_{\bar{c}}+a \tilde{\partial}_{\bar{c}} )h^{a\bar{c}} 
\bigg\}\;. 
\end{split}
\end{equation}
Following the above procedure one obtains the decoupled  quadratic action in terms of gauge invariant variables: 
\begin{equation}
\begin{split}
\label{GI ACTION BEFORE HAMIL}
S= \int & dt  \int d^{2d}{\bf X} \,  n\, e^{-2\Phi}
\big( \mathcal{L}_T + \mathcal{L}_V + \mathcal{L}_S \big) \,, 
\end{split}
\end{equation}
where  $\mathcal{L}_T$, $\mathcal{L}_V$ and  $\mathcal{L}_S$ denote the Lagrangians 
for the tensor, vector and scalar modes, respectively. As emphasized before,  the action 
should completely decouple among these modes,  as indeed it does. Specifically, 
the action for the tensor modes is given by 
\begin{equation}
\mathcal{L}_T \ = \   \dot{\widehat{h}}_{a\bar{b}} \,\dot{\widehat{h}}^{a\bar{b}}   + 2 H^2\, \widehat{h}_{a\bar{b}}\, \widehat{h}^{a\bar{b} }-  2 H^2 g_{a}{}^{\bar{c}}g_{\bar{d}}{}^b  \,\widehat{h}^{a\bar{d}}\, \widehat{h}_{b\bar{c}} + \widehat{h}_{a\bar{b}} \Delta  \widehat{h}^{a\bar{b}}\,, 
\end{equation}
while the action for the vector modes reads 
\begin{equation}
\begin{split}
\mathcal{L}_V  \, = \, & - \dot{\widehat{B}}_{\bar{a}}\Delta \dot{\widehat{B}}^{\bar{a}} - \dot{\widehat{B}}^a\Delta\dot{\widehat{B}}_a -  H^2 \widehat{B}_a\Delta \widehat{B}^a - H^2 \widehat{B}_{\bar{a}} \Delta \widehat{B}^{\bar{a}}  
\\& + 4H^2 g_{a}{}^{\bar{b}} \widehat{B}^a \Diamond \widehat{B}_{\bar{b}} 
   - \widehat{B}_a (\Delta^2-\Diamond^2) \widehat{B}^a
   -\widehat{B}_{\bar{a}} (\Delta^2-\Diamond^2) \widehat{B}^{\bar{a}} 
\\&
- \frac{1}{2} \widehat{A}_{\bar{a}} \Delta \widehat{A}^{\bar{a}} - \frac{1}{2} \widehat{A}_a \Delta  \widehat{A}^a 
 - \sqrt{2} \widehat{A}^a \Diamond(  \dot{\widehat{B}}_a  +Hg_{a}{}^{\bar{b}} \widehat{B}_{\bar{b}} )
+ \sqrt{2} \widehat{A}^{\bar{a}} \Diamond (\dot{\widehat{B}}_{\bar{a}}+Hg_{\bar{a}}{}^{b}\widehat{B}_b)    \,. 
\end{split}
\end{equation}
Note that while this action carries two more vector modes than in standard gravity, the $A$ modes are actually 
auxiliary as they may be eliminated by their own equations of motions. Indeed, 
varying with respect to $\widehat{A}_a$ and $\widehat{A}_{\bar{a}}$ and assuming $\Delta$ is invertible, we obtain 
\begin{equation}
\begin{split}
\widehat{A}_a & = -\sqrt{2} \Delta^{-1} \Diamond ( \dot{\widehat{B}}_a + Hg_{a}{}^{\bar{b}} \widehat{B}_{\bar{b}})    \,,\\
\widehat{A}_{\bar{a}} & = \sqrt{2} \Delta^{-1} \Diamond ( \dot{\widehat{B}}_{\bar{a}} +Hg_{\bar{a}}{}^b  \widehat{B}_b )  
 \,, 
\end{split}
\end{equation}
which may be reinserted into the action to obtain: 
 \be
 \begin{split}
  {\cal L}_V(B) = &  - \dot{\widehat{B}}_{\bar{a}}\Delta \dot{\widehat{B}}^{\bar{a}} - \dot{\widehat{B}}^a\Delta\dot{\widehat{B}}_a - H^2 \widehat{B}_a\Delta \widehat{B}^a -H^2 \widehat{B}_{\bar{a}} \Delta \widehat{B}^{\bar{a}}  \\& 
  +4H^2 g_{\bar{b}}{}^a \widehat{B}^{\bar{b}}\Diamond \widehat{B}_{a} 
   -\widehat{B}_{\bar{a}} (\Delta^2-\Diamond^2) \widehat{B}^{\bar{a}} - \widehat{B}_a (\Delta^2 -\Diamond^2)\widehat{B}^a\\&
  + ( \dot{\widehat{B}}_{\bar{a}} + H g_{\bar{a}}{}^b \widehat{B}_b) \Delta^{-1} \Diamond^2 ( \dot{\widehat{B}}^{\bar{a}} + H g^{\bar{a}c}\widehat{B}_c) + ( \dot{\widehat{B}}_a + Hg_{a}{}^{\bar{b}} \widehat{B}_{\bar{b}}) \Delta^{-1} \Diamond^2 ( \dot{\widehat{B}}^a + Hg^{a\bar{c}}\widehat{B}_{\bar{c}})  \,. 
 \end{split}
 \ee
The action for the scalar modes is
\begin{equation}
  \begin{split}
 \mathcal{L}_S  = & - 4\dot{\widehat{\varphi}}^2 + 8 \dot{\Phi}\phi\dot{\widehat{\varphi}} + d\dot{\widehat{E}}^2 + 2dH ( \widehat{\phi} + 2 \widehat{\varphi}) \dot{\widehat{E}} 
  + \frac{1}{4} \dot{\widehat{C}} ( \Delta^2 - \Diamond^2 )\dot{\widehat{C}}  + \frac{1}{2} H^2 \widehat{C}( \Delta^2 - \Diamond^2 )\widehat{C}
  \\&
 + \frac{\sqrt{2}}{2}\widehat{A} \Diamond (2 \dot{\widehat{\varphi}} - 2\dot{\Phi} \widehat{\phi} +  d H \widehat{E} ) + \frac{\sqrt{2}}{2}   \widehat{A} \Delta \big( H(\widehat{\phi} + 2 \widehat{\varphi}) +  \dot{\widehat{E}}\big ) \\&
-  \frac{\sqrt{2}}{4} H \widehat{A} ( \Delta^2 - \Diamond^2) \widehat{C}   + \frac{1}{16} \widehat{A} ( \Delta^2 - \Diamond^2) \widehat{A} - \frac{\sqrt{2}}{8} \widehat{\bar{A}} ( \Delta^2 - \Diamond^2) \dot{\widehat{C}}   + \frac{1}{16} \widehat{\bar{A}} ( \Delta^2 - \Diamond^2)   \widehat{\bar{A}} 
  \\&
 + 4\widehat{\phi} \Delta \widehat{\varphi} - 4 \widehat{\varphi }\Delta \widehat{ \varphi} + ( d-2) \widehat{E} \Delta \widehat{E} + 2 ( \widehat{\phi} - 2 \widehat{\varphi}) \Diamond\widehat{ E} - \frac{1}{2} ( \widehat{\phi} -2 \widehat{\varphi}) (\Delta^2 -\Diamond^2) \widehat{C}\,. 
 \end{split}
 \end{equation}
Here, $\widehat{\bar{A}}$, and $\widehat{A}$ are auxiliary fields. Varying the action with respect to these fields yields 
\begin{equation}
( \Delta^2 - \Diamond^2) (-\sqrt{2} \dot{\widehat{C}} + \widehat{\bar{A}}  ) = 0  \,, 
\end{equation}
 \begin{equation}
\frac{1}{8} ( \Delta^2 - \Diamond^2) (\widehat{A}-2\sqrt{2} H \widehat{C}) + \frac{\sqrt{2}}{2} \Diamond ( 2 \dot{\widehat{\varphi}}   - 2 \dot{\Phi} \widehat{\phi} + d H \widehat{E} ) + \frac{\sqrt{2}}{2}   \Delta \big( H( \widehat{\phi} + 2 \widehat{\varphi}) + \dot{\widehat{E}} \big)  = 0 \,. 
\end{equation}
Since $\Delta^2 - \Diamond^2 = 4 \partial^2 \tilde{\partial}^2$ and $\partial^2$ and $\tilde{\partial}^2$ are invertible, we can solve for $\widehat{\bar{A}}$ and $\widehat{A}$, 
\begin{flalign}
\widehat{\bar{A}} &= \sqrt{2} \dot{\widehat{C}} \,,\\ 
\widehat{A} &= 2 \sqrt{2} H \widehat{C} - \sqrt{2} \partial^{-2}\tilde{\partial}^{-2} \big( \Diamond ( 2 \dot{\widehat{ \varphi}}   - 2 \dot{\Phi} \widehat{ \phi} + d H \widehat{ E} )  +  \Delta \big(H(\widehat{ \phi} + 2 \widehat{ \varphi}) + \dot{\widehat{E}} \big)   \,, 
\end{flalign} 
and re-insert these into the action, which yields 
\begin{equation}
  \begin{split}
 \mathcal{L}_S  =& 
 \ ( d-1)\dot{\widehat{E}}^2 + 2(d-1)H ( \widehat{\phi} + 2 \widehat{\varphi}) \dot{\widehat{E}}  
  + \frac{1}{8} \dot{\widehat{C}} ( \Delta^2 - \Diamond^2 )\dot{\widehat{C}}   
  \\&
 +2 \widehat{C} \Diamond (2 \dot{\widehat{\varphi}} - 2\dot{\Phi} \widehat{\phi} +  d H \widehat{E} ) + 2  \widehat{C}\Delta  \big(H(\widehat{ \phi} + 2 \widehat{ \varphi}) + \dot{\widehat{E}} \big)  
  \\&
  +
  ( d - 1)H^2 \widehat{\phi}^2 -4 H^2 \widehat{\phi} \widehat{\varphi} - 4 H^2 \widehat{\varphi}^2 + d^2 H^2 \widehat{E}^2+ 4 d H \dot{\widehat{\varphi} }\widehat{E}  - 4 d H \dot{\Phi} \widehat{E} \widehat{\phi} \\&
   + 4\widehat{\phi} \Delta \widehat{\varphi} - 4 \widehat{\varphi }\Delta \widehat{ \varphi} + ( d-2) \widehat{E} \Delta \widehat{E} + 2 ( \widehat{\phi} - 2 \widehat{\varphi}) \Diamond\widehat{ E} - \frac{1}{2} ( \widehat{\phi} -2 \widehat{\varphi}) (\Delta^2 -\Diamond^2) \widehat{C}
  \\&
 - \frac{1}{2} ( 2 \dot{\widehat{\varphi}} - 2 \dot{\Phi} \widehat{\phi} + d H \widehat{E} )\partial^{-2}\tilde{\partial}^{-2} ( a^4 \tilde{\partial}^4 + a^{-4} \partial^4) ( 2 \dot{\widehat{\varphi}} - 2 \dot{\Phi} \widehat{\phi} + d H \widehat{E} )\\& 
 - \frac{1}{2} \big( H ( \widehat{\phi} + 2 \widehat{\varphi}) + \dot{\widehat{E}} \big) \partial^{-2}\tilde{\partial}^{-2} ( a^4 \tilde{\partial}^4 + a^{-4} \partial^4)  \big( H ( \widehat{\phi} + 2 \widehat{\varphi}) + \dot{\widehat{E}} \big)  \\&
 - ( 2 \dot{\widehat{\varphi}} - 2 \dot{\Phi} \widehat{\phi} + d H \widehat{E} ) \partial^{-2}\tilde{\partial}^{-2} ( a^4 \tilde{\partial}^4 -  a^{-4} \partial^4) \big( H ( \widehat{\phi} + 2\widehat{ \varphi}) + \dot{\widehat{E}} \big) \,.
 \end{split}
 \end{equation}

We close this section by pointing out that the definition of gauge invariant variables can be extended to cubic order, 
using the results of \cite{Chiaffrino:2020akd}. This approach uses the insight that the passing over to gauge invariant variables 
has a natural interpretation in the framework of $L_{\infty}$-algebras as homotopy transfer, which 
in turn yields an algorithmic procedure to extend the gauge invariant functions to higher order. As before, the cubic 
action in terms of these variables takes the same form as the gauge fixed action, but with the fields being the fully gauge invariant ones. 
We will leave the details to future work.

\section{Summary and Outlook}

In this paper we have analyzed general aspects of a genuine double field theory on time-dependent (cosmological) 
backgrounds with zero spatial curvature, with a particular emphasis on the presence of winding modes when 
the spatial topology is a torus. Indeed, to quadratic and cubic order  the theory is 
fully consistent in a weakly constrained sense, meaning that the fields genuinely depend on doubled coordinates, 
hence encoding both momentum and winding modes. We presented a self-contained discussion of the resolution 
of the consistency issues related to the weak constraint, 
including the role of cocycle factors. Finally, we analyzed, for FRW backgrounds 
with a single scale factor $a(t)$,  the independent gauge invariant 
modes under a scalar-vector-tensor decomposition. This analysis is significantly more involved compared 
to standard cosmology, due to the doubled derivatives, and also shows some striking new features. 
Most importantly, the nature of the gauge invariant modes changes relative to standard cosmology  in that there are more vector and scalar modes, 
while the number of tensor modes is reduced (leaving, of course, the total number of off-shell degrees of freedom 
unchanged). 

In this paper we have just taken the first steps in developing a full-fledged cosmological 
perturbation theory for genuine double field theory. It remains to extend this framework in a way 
that allows for (semi-)realistic cosmological scenarios, so that a confrontation with observation 
may eventually become feasible. Concretely, the following research problems present themselves:

\begin{itemize}

\item In this paper we have not included matter other than the scalar dilaton and Kalb-Ramond B-field 
that are automatically encoded in double field theory. The latter fields are insufficient in order to provide semi-realistic 
cosmologies, so it is important to add further matter, perhaps in the form of a generic  
duality invariant energy-momentum tensor.

\item In string theory there is an infinite number of higher-derivative $\alpha'$ corrections, 
which plausibly are relevant in the very early universe but which have not been included here. 
While the complete $\alpha'$ corrections
remain of course unknown, remarkably, it is possible to give a complete classification of all corrections 
compatible with $O(d,d)$ duality invariance for purely time-dependent cosmological backgrounds \cite{Hohm:2015doa}. 
In order to set up the corresponding $\alpha'$-complete cosmological perturbation theory 
one might hope that for the quadratic approximation one can bypass the hard problem 
of defining the full double field theory with all $\alpha'$ corrections included.  
Assuming  (as on flat space \cite{Zwiebach:1985uq}) that there is a field basis  for which the
quadratic terms are two-derivative one may try to write directly
a quadratic action that is gauge invariant, but using now the known $\alpha'$-completed background Friedmann equations, 
which include only first-order time derivatives \cite{Hohm:2015doa}.

\item Recently, an intriguingly simple off-shell and gauge invariant double copy relation has been uncovered between 
Yang-Mills theory and double field theory on flat space  \cite{Diaz-Jaramillo:2021wtl}, at least to 
cubic order in fields. The natural question arises whether there are similar double copy relations for 
double field theory on general backgrounds or at least on the purely time-dependent backgrounds explored here.

\item Arguably the most important open research project  is the computation of cosmological 
correlation functions, as these may eventually relate to the CMB and hence allow one to discriminate 
between different cosmological scenarios. In particular, since the independent gauge invariant modes have a 
different character in a genuine double field theory, meaning the degrees of freedom are distributed differently 
among scalar, vector and tensor modes,  it seems quite promising that we will eventually be able to 
distinguish it observationally from general relativity coupled to ordinary matter. 
Since cosmological perturbation theory in general, and the computation of cosmological correlation functions 
in particular, is infamously opaque \cite{Mukhanov:1990me,Maldacena:2002vr}, to this end 
it would be important to develop more systematic techniques. We believe that a most promising framework is that 
of the homotopy algebra formulation of gauge theories, see e.g.~\cite{Hohm:2017pnh}, 
because here there are natural `homological' approaches to the definition of gauge invariant variables 
\cite{Chiaffrino:2020akd} and to the computation of quantum mechanical expectation values \cite{Chiaffrino:2021pob}.

\end{itemize}

\subsection*{Acknowledgements} 
We are grateful to Heliudson Bernardo, Roberto Bonezzi, 
Robert Brandenberger,  Christoph Chiaffrino, Chris Hull, Matt Kleban, 
Jean-Luc Lehners and Barton Zwiebach for useful discussions, correspondence and 
related collaborations.

This work is funded   by the European Research Council (ERC) under the European Union's Horizon 2020 research and innovation programme (grant agreement No 771862)
and by the Deutsche Forschungsgemeinschaft (DFG, German Research Foundation), 
``Rethinking Quantum Field Theory", Projektnummer 417533893/GRK2575.

\end{document}